\documentclass[a4paper,fleqn]{cas-dc}
\usepackage[numbers]{natbib}
\usepackage{graphicx}
\usepackage{subcaption}
\usepackage{caption}
\usepackage{lastpage}
\usepackage{fancyhdr}

\def\tsc#1{\csdef{#1}{\textsc{\lowercase{#1}}\xspace}}
\tsc{WGM}
\tsc{QE}

\begin{document}
\let\WriteBookmarks\relax
\def\floatpagepagefraction{1}
\def\textpagefraction{.001}

\shorttitle{} 

\shortauthors{Shibing Xiang et~al.}
\title [mode = title]{SciConNav: Knowledge navigation through contextual learning of extensive scientific research trajectories}
\author[1,2]{Shibing Xiang}
\author[3,4]{Xin Jiang}
\author[5,6]{Bing Liu}
\author[1]{Yurui Huang}
\author[1]{Chaolin Tian}

\author[1]{Yifang Ma}[orcid=0000-0003-0326-7993]
\ead{mayf@sustech.edu.cn}
\cormark[1] 
\cortext[1]{Corresponding author}

\affiliation[1]{organization={Department of Statistics and Data Science, Southern University of Science and Technology},
  city={Shenzhen},
  state={Guandong},
  country={China}}
\affiliation[2]{organization={Peng Cheng Laboratory},
  city={Shenzhen},
  state={Guandong},
  country={China}}
\affiliation[3]{organization={Zhongguancun Laboratory},
  city={Beijing},
  country={China}}
\affiliation[4]{organization={NLSDE \& Institute of Artificial Intelligence, Beihang University},
  city={Beijing},
  country={China}}
\affiliation[5]{organization={LMIB \& School of Mathematical Sciences, Beihang University},
  city={Beijing},
  country={China}}
\affiliation[6]{organization={Zhengzhou Aerotropolis Institute of Artificial Intelligence},
  city={Zhengzhou},
  state={Henan},
  country={China}}

\begin{abstract}
New knowledge builds upon existing foundations, which means an interdependent relationship exists between knowledge, manifested in the historical development of the scientific system for hundreds of years. By leveraging natural language processing techniques, this study introduces the Scientific Concept Navigator (SciConNav), an embedding-based navigation model to infer the ``knowledge pathway'' from the research trajectories of millions of scholars. We validate that the learned representations effectively delineate disciplinary boundaries and capture the intricate relationships between diverse concepts. The utility of the inferred navigation space is showcased through multiple applications. Firstly, we demonstrated the multi-step analogy inferences within the knowledge space and the interconnectivity between concepts in different disciplines. Secondly, we formulated the attribute dimensions of knowledge across domains, observing the distributional shifts in the arrangement of 19 disciplines along these conceptual dimensions, including ``Theoretical'' to ``Applied'', and ``Chemical'' to ``Biomedical', highlighting the evolution of functional attributes within knowledge domains. Lastly, by analyzing the high-dimensional knowledge network structure, we found that knowledge connects with shorter global pathways, and interdisciplinary knowledge plays a critical role in the accessibility of the global knowledge network. Our framework offers a novel approach to mining knowledge inheritance pathways in extensive scientific literature, which is of great significance for understanding scientific progression patterns, tailoring scientific learning trajectories, and accelerating scientific progress. 
\end{abstract}

\begin{keywords}
Knowledge Navigation\sep Neural Embedding\sep Research Trajectories\sep  Science of Science\sep Computational Social Science
\end{keywords}

\begin{highlights}
    \item We learned the embedding space for knowledge navigation based on the research trajectories of millions of scholars, and proposed the Scientific Concept Navigator (SciConNav) to navigate knowledge inheritance pathways.
    \item We analyzed and verified the effectiveness of the learned representations by aligning them with external concept trees by cosine similarity.
    \item We demonstrated the utility of SciConNav for multi-step analogies of new concepts and knowledge navigation, enhancing decision-making and learning guidance by identifying highly interconnected concepts step-by-step.
    \item We identified key concepts by assessing the accessibilities, highlighting the critical bridging role of interdisciplinary concepts in the global knowledge network, thus facilitating efficient navigation through complex scientific information.
\end{highlights}



\maketitle

\section{Introduction}\label{introduction_section}

In the current science and technology landscape, we are experiencing an unprecedented expansion of knowledge and information overload. Knowledge manifests in various tangible forms, including reports, papers, books, mathematical formulas, and diagrams\cite{abdullah2002knowledge}. This growth is evident in the increasing number of publications, research topics, and patents, all documented in academic databases and extensive scientific knowledge graphs like Wikipedia, Web of Science\cite{xu2023exploring}, Scopus, PubMed\cite{white2020pubmed}, Dimensions\cite{hook2018dimensions}, OpenAlex\cite{priem2022openalex}, and Aminer\cite{tang2016aminer}.

Researchers constantly face the challenge of selecting the next topic for exploration. The overwhelming influx of research and information makes it difficult for scientists to stay abreast with the latest developments, adopt interdisciplinary approaches, and identify the most rational research paths to scientific discoveries and advanced technologies\cite{shirah2023computer}. This challenge is particularly daunting for early-career scientists who may not yet have a broad knowledge base in their respective fields. Knowledge navigation\cite{hammond1995faq,hammond1995case,li2014recommender} has emerged as a viable solution, serving as intelligent assistance for navigation\cite{amant1998experimental}, providing access to valuable and reliable information\cite{jadad1998rating}, assisting learners and researchers in navigating through the expanding information spaces\cite{burke1996knowledge,benyon2001new} and the complexities of modern scientific research. Understanding complex knowledge structures and their dependency relationships is crucial for effectively navigating the intellectually structured vast knowledge landscape and providing access to valuable and reliable information\cite{jadad1998rating}.

Previous researches on knowledge navigation emphasize computer-based systems or platforms for information retrieval and well-organized knowledge repositories with complex interactions\cite{patel1998understanding} to facilitate easy access and intuitive navigation. For instance, the McSyBi\cite{yamamoto2007biomedical} navigation system assists in acquiring knowledge from biomedical literature, offering researchers a comprehensive overview of topics and their interrelationships. The TaxoFolk\cite{kiu2010taxofolk} introduced hybrid taxonomy-folksonomy classifications to enhance knowledge retrieval. Similarly, CoNavigator\cite{hao2021conavigator} employs formal concept analysis for domain-specific knowledge retrieval, particularly in the context of COVID-19. The Knowledge Navigator Model (KNM)\cite{hsieh2009construction,hsieh2020evolution} guides structured knowledge management within organizations, representing high-level behavioral or implementation navigation. Information retrieval-based knowledge navigation helps users acquire meta-knowledge. However, these methods seldom incorporate representation learning techniques to capture contextual semantics in natural language processing (NLP), fail to utilize the interdependencies between knowledge for more accurate retrieval, and often fall short in identifying highly relevant global pathways to desired knowledge.

In the field of the Science of Science\cite{zeng2017science,fortunato2018science}, research has primarily focused on knowledge mapping\cite{vail1999knowledge}, which involves formulating the graphical knowledge map to analyze the latent structure of knowledge\cite{chiu2014topic,miao2022latent} and learning-dependency\cite{liu2012topological}, assist users in navigating complex issues or information\cite {wexler2001and}. Mapping refers to the process of creating a visual representation of information or knowledge in this context\cite{wexler2001and}. Knowledge mapping\cite{wexler2001and} facilitates the development of knowledge management by utilizing graphical representations to display knowledge entities and their interrelationships, such as the construction of co-occurrence or collaboration networks\cite{he2022knowledge}, extraction of impactful authors, articles, and affiliations\cite{ding2022knowledge}, as well as analyzing the current research status and predicting future topics\cite{dwivedi2023evolution}. Visualization of knowledge maps\cite{borner2003visualizing,shiffrin2004mapping}, with leading software such as CiteSpace\cite{chen2006citespace} and VOSViewer\cite{van2010software} continuously evolving, is crucial in this field. Knowledge mapping methods construct the conceptual knowledge map and reveal extensive information from academic entities. Recent studies aim to uncover the latent knowledge landscape\cite{gonzalez2023landscape} using embedding techniques in NLP. Notable efforts involve creating embedding atlases for journals\cite{peng2021neural}, papers\cite{ganguly2017paper2vec}, and concepts\cite{choi2016multi,chen2020bioconceptvec}, as well as tracing the trajectories of research affiliations\cite{murray2023unsupervised}.

These efforts aim to explore, manage, and share explicit knowledge effectively. However, these constructed graphical knowledge maps often overlook knowledge precedence relations and intrinsic dependences. In most cases, representations of scientific entities in embedding space are primarily used for macro-level visualization and are rarely utilized for knowledge navigation, especially at the level of hierarchical scientific concepts. Furthermore, these embedding methods fail to leverage prior knowledge and prerequisite relations\cite{scheines2014discovering,manrique2019towards}, or synonymously, the precedence relations\cite{manrique2018investigating,xiao2021mining} encoded in historical records of scientific publications. 
Given the abundance of scientific information available, navigating an efficient, logical, and reasonable pathway on the embedding map for learning purposes and acquiring knowledge is essential. 

To address the shortcomings of the aforementioned research, we leverage the conceptual knowledge embedding map for knowledge navigation. Our study addresses the following questions: (1) Topic Selection: How can researchers choose a topic closest to their background if they wish to switch to another field? (2) Interdependent Learning Pathways: How can researchers identify the interdependent learning pathways after selecting a topic from a different domain or exploring a topic of interest in various disciplines? 

In response to these challenges, we utilized the concept entity in the large-scale scientific corpus OpenAlex\cite{priem2022openalex} as metaknowledge. We developed the Scientific Concept Navigator (SciConNav) to address the growing need for fine-grained knowledge acquisition. SciConNav regards experienced scientists as comprehensive mentors rich in prior knowledge and leverage their extensive research trajectories as a training corpus to learn the concept representations. The embedding representations\cite{li2013construction} underpinning the global knowledge navigation space enable SciConNav to provide effective research or curriculum planning, learning guidance\cite{chiou2010adaptive,diaz2024artificial}, precise knowledge exploration and discovery\cite{yan2024impact}, and science and technology forecasting.

The SciConNav model leverages the NLP embedding technique to learn the contextual representation of concepts and conduct knowledge navigation within hierarchical scientific concepts. The learned representations enable the identification of related topics based on existing knowledge and the inference of highly interconnected concepts in a step-by-step learning pathway. This pathway encodes the interdependence and temporal order of prerequisites between concepts across diverse disciplines. This research serves as an initial exploration into understanding the interconnected landscape of concepts, empowering researchers to acquire the desired knowledge swiftly. The embedded knowledge space provides insights for navigating logical learning paths, establishing connections between new knowledge through the shortest path, and uncovering interconnected entities efficiently. Our key contributions are:
\begin{enumerate}
    \item We learn the embedding space for knowledge navigation based on the research trajectories filled with concepts of millions of scholars. 
    \item We demonstrated the utility of the embedding space for multi-step analogy inference of new concepts to enhance decision-making and creative thinking.
    \item We introduced \textbf{SciConNav} as a novel approach to navigating the knowledge inheritance pathways learned from extensive scientific research trajectories. 
    \item We identified key concepts using centrality measures, highlighting the critical bridging role of interdisciplinary concepts in the global knowledge network. 
\end{enumerate}

\begin{figure*}
    \centering
    \includegraphics[width=0.9\textwidth]{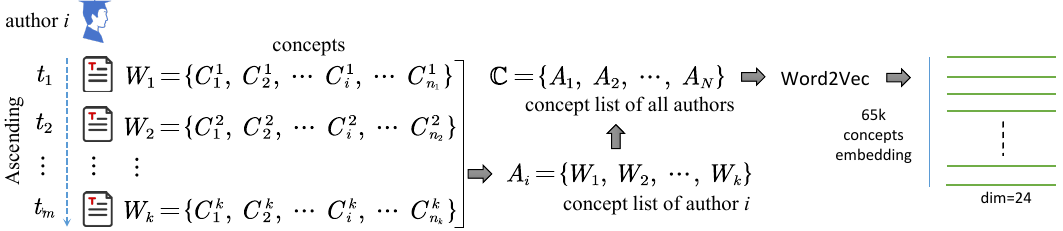}
    \caption{\textbf{Concept representation learning procedure from extensive research trajectories}. For each author, the research concepts of each of her/his papers are sorted by the publication years. Then, we concatenate the concept list of all selected authors and obtain embedding vectors of all concepts by training concatenated concept lists with the Word2Vec model.}
    \label{concept_representation_learning}
\end{figure*}

\begin{table*}[h]
    \centering
    \caption{Illustration of concept discipline category classification, note that $S\cup M = S^+ \cup M^-$.}
    \begin{tabular}{l|cccccccc}
    \hline
    Type  & \multicolumn{2}{c|}{$S$ (single root)} & \multicolumn{6}{c}{$M$ (mutiple roots / interdisciplinary)}    \\ \hline
    Concept & \multicolumn{2}{c|}{Simplicial manifold} & \multicolumn{2}{c|}{PPADS} & \multicolumn{2}{c|}{Glycine cleavage system} & \multicolumn{2}{c}{Cutter location} \\ \hline
    \multirow{5}{*}{Ancestor}      & Root  & \multicolumn{1}{c|}{\# paths} & Root  & \multicolumn{1}{c|}{\# paths} & Root  & \multicolumn{1}{c|}{\# paths} & Root     & \# paths      \\
    & \textcolor[rgb]{0,0,1}{\textbf{Math.}} & \multicolumn{1}{c|}{16}  & \textcolor[rgb]{0,0,1}{\textbf{Bio.}}  & \multicolumn{1}{c|}{54}  & \textcolor[rgb]{0,0,1}{\textbf{Bio.}}  & \multicolumn{1}{c|}{29}  & \textcolor[rgb]{0,0,1}{\textbf{C.S.}}     & 12     \\
    &  & \multicolumn{1}{c|}{}  & Chem. & \multicolumn{1}{c|}{32}  & Chem. & \multicolumn{1}{c|}{22}  & \textcolor[rgb]{0,0,1}{\textbf{Eng.}}     & 12     \\
    &  & \multicolumn{1}{c|}{}  & Med.  & \multicolumn{1}{c|}{37}  & Phys. & \multicolumn{1}{c|}{8}   & \textcolor[rgb]{0,0,1}{\textbf{Mat.Sci.}}      & 12     \\
    &  & \multicolumn{1}{c|}{}  &  & \multicolumn{1}{c|}{}  &  & \multicolumn{1}{c|}{}  & Math.    & 4      \\ \hline
    \multirow{2}{*}{Classification} & \multicolumn{2}{c|}{Mathematics}      & \multicolumn{2}{c|}{Biology}   & \multicolumn{2}{c|}{Biology}   & \multicolumn{2}{c}{Multi-interdisciplinary} \\ \cline{2-9} 
    & \multicolumn{6}{c|}{classifiable / distinguishable}  & \multicolumn{2}{c}{indistinguishable}  \\ \hline
    Updated    & \multicolumn{6}{c|}{$S^+$ (Disciplinary)}     & \multicolumn{2}{c}{$M^-$ (Multi-interdisciplinary)}      \\ \hline
    \end{tabular}
    \label{discipline_classification_example}
\end{table*}

\section{Methods and conceptual validation}\label{methods_section}
We constructed the concept sequences for each researcher, further trained the embedding representations with curated concept sequences of millions of authors, and validated the effectiveness and similarity properties using an external OpenAlex concept tree. Most importantly, we analyze the roles and variance of attributes based on the specifically designed conceptual dimension, which helps us understand the semantic functionality of knowledge from different disciplines.

\subsection{Knowledge entity and discipline classification}
\begin{figure*}[H]
    \centering
    \begin{subfigure}[c]{0.18\textwidth}
  \includegraphics[width=\linewidth]{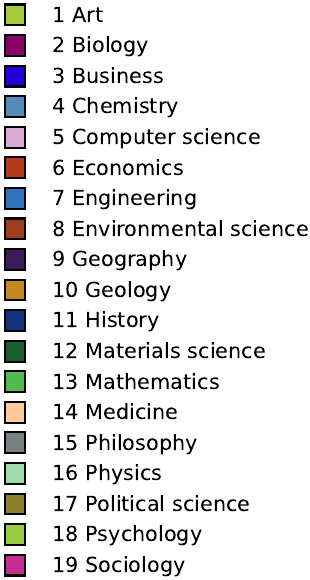}
    \end{subfigure}
    \hfill
    \begin{subfigure}[c]{0.4\textwidth}
        \includegraphics[width=\linewidth]{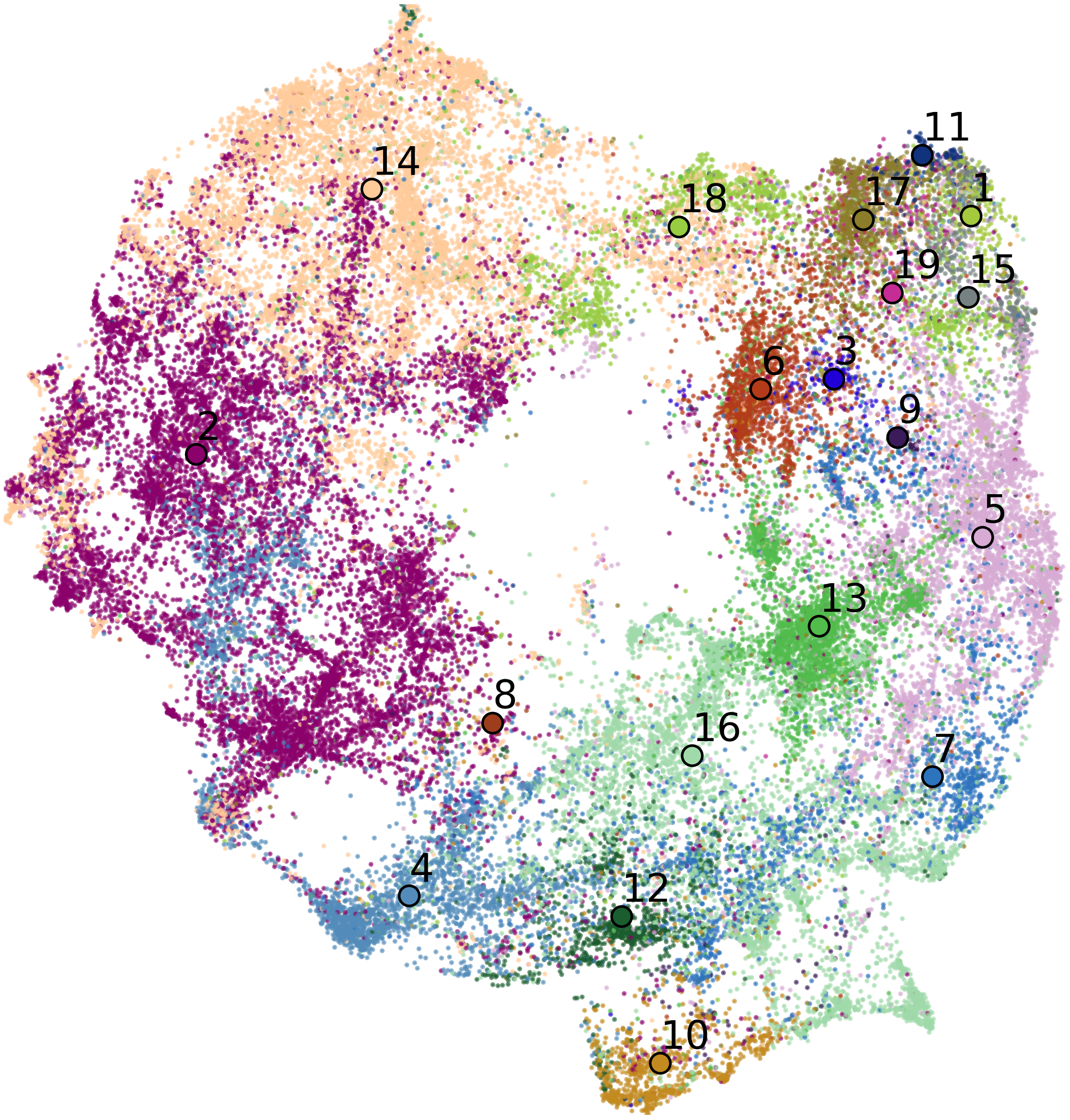}
        \caption{Concept of 19 distinct disciplines}
        \label{figure_2_19_disciplines}
    \end{subfigure}
    \hfill
    \begin{subfigure}[c]{0.4\textwidth}
        \includegraphics[width=\linewidth]{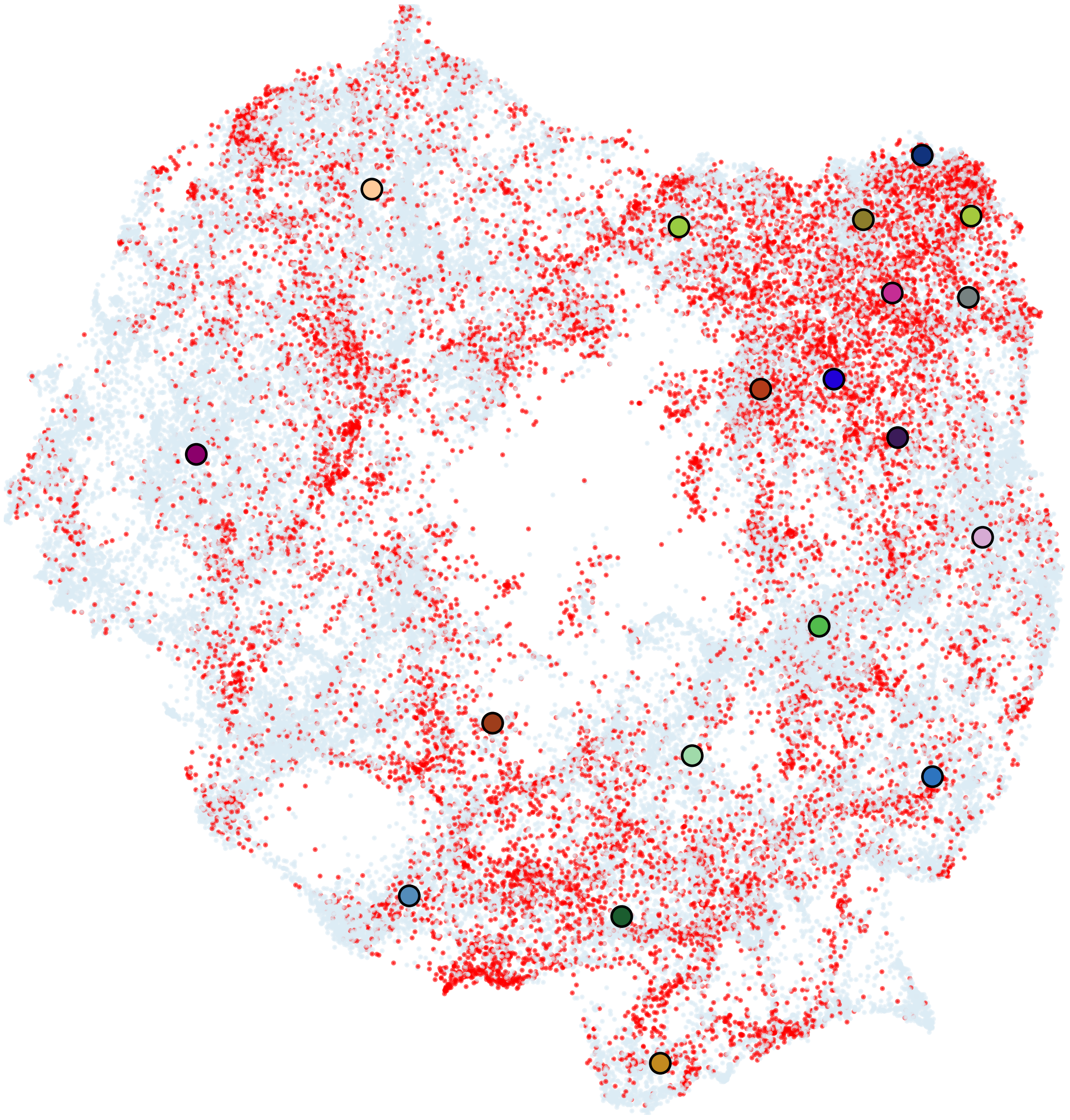}
        \caption{Comparison with multi-disciplinary}
        \label{figure_2_multi}
    \end{subfigure}
 \caption{\textbf{UMAP dimensionality reduction of concept representation}. For each color-coded set of dots, the point with the highest density is circled in black, and the adjacent number indicates the corresponding discipline shown on the left side of the diagram. (a) Embedding map of concepts from all disciplines with UMAP 2D dimension reduction, concepts with the same color belong to the same discipline, while each discipline is represented and sorted by a number from 1 to 19. (b) Overview of ``Multi-interdisciplinary'' concepts (red) and comparison with other ``disciplinary'' concepts (cyan). } 
    \label{figure_2}
\end{figure*}

The OpenAlex dataset, which is publicly accessible, employs a hierarchical tree structure to systematically organize concepts and knowledge. Each concept corresponds to a Wikidata item. This structure comprises 19 root-level concepts (level 0) representing 19 disciplines, as shown on the left side of Fig. \ref{figure_2_19_disciplines}. The remaining concepts are descendants spanning five levels (level 1 to level 5) below the root concepts. For a detailed hierarchical sub-concept tree structure of Mathematics, Computer Science, Physics, and Biology, refer to Fig. \ref{examples_of_sub_concept_tree} in the Supplementary Information (SI).

In this study, we utilized concepts as knowledge entities. Initially, there were 65,068 concepts. After removing 471 concepts with no ancestor root and concepts of low frequency, the dataset comprised 64,510 concepts. For any remaining concept $c$, we define its ancestor roots (AR) of $c$ as the set of disciplines that are root-level ancestors of $c$. Conversely, non-ancestor roots (NAR) are disciplines that do not directly connect to $c$. Thus, NAR is the complement of AR. For example, consider the concept ``PPADS'' in   Table \ref{discipline_classification_example}, its AR = $\left\{\text{Biology}, \text{Chemistry}, \text{Medicine}\right\}$. We then classified these concepts into disciplines using a two-step approach. Firstly, we traced the lineage of concept $c$ back to the root-level concepts, identifying all potential AR. For instance, a discipline $d$ is considered the ancestor of concept $c$ if at least one graph path connects $d$ and $c$, the example of concept trees linking $d$ and $c$ can be found in Fig. \ref{examples_of_sub_concept_tree} in SI. Secondly, we assigned each concept $c$ to a discipline category based on its highest degree of membership to these identified AR. The membership degree of concept $c$ to discipline $d$ is determined by the number of graph paths between them. A detailed illustration of concept discipline category classification can be seen in  Table \ref{discipline_classification_example}, the abbreviation of AR with the highest number of paths (\# paths) are marked with blue color and bold font.  

We denote $S$ as the set of concepts that has only one AR and $M$ as the concept set that has multiple AR ($\geqslant 2$), and we define the concept $c\in M$ as ``interdisciplinary'' concepts. As concept $c\in S$ has only one AR, then this single AR $d$ is determined as the discipline category of concept $c$ ($c=$ Simplicial manifold, $d=$ Mathematics). If concept $c\in M$, we determine the discipline category of ``interdisciplinary'' concept $c$ as the root concept that has the maximal number of graph paths to the concept $c$ ($c=$ PPADS, $d=$ Biology), thus a subset of ``interdisciplinary'' concepts in $M$ are classifiable, and we define the set of all classifiable concepts as $S^+$.
Furtherly, if there are multiple roots with the same maximal number of paths to concept $c$, we tag such ``interdisciplinary'' concept $c$ as ``Multi-interdisciplinary.'' ($c=$ Cutter location, $d=$ ``Multi-interdisciplinary''). Let $M^-$ denote the set of all concepts labeled ``Multi-interdisciplinary'' (a total of 15,261 concepts, with multiple indistinguishable root disciplines). In comparison with ``Multi-interdisciplinary'' in $M^-$, we refer to all concepts that are classifiable in $S^+$ as ``Disciplinary'' concepts and note that $S\cup M = S^+ \cup M^-$.

\subsection{Extraction and learning of research trajectory}
In academic research advancement, scientists often conduct new research topics based on their previous background, which implies the underlying interdependent relationship, intuitively the prerequisite, or the precedence relationship\cite{lu2019concept}, between the previous topics and the new topics. The prerequisite relationships\cite{tang2023continual} are foundations for knowledge space theory\cite{doignon2015knowledge,wang2023new} and knowledge tracing\cite{gan2022knowledge,lu2022cmkt}. The precedence relationship among concepts plays an important role in applications like personalized learning\cite{feldman2024navigating} and teaching\cite{chen2018prerequisite}, designing instructional rules for courses\cite{gasparetti2015exploiting}, knowledge recommendation\cite{he2022exercise}, curriculum planning\cite{sun2022conlearn}, etc. Thus, we leverage the prerequisite relations\cite{tang2023continual} among scientific concepts, essentially the temporal order between knowledge entities encoded in research trajectories of millions of scholars, which are also the key to enhancing the interpretability of knowledge causality. 

Given that prior knowledge and strategies significantly impact task completion during navigation\cite{fuentes2000internet}, and historical records are important prior for patterns mining\cite{guerbas2013effective}. We constructed the research trajectories of scholars based on their historical publication records. The research trajectory with concepts can provide context enriched with semantic information and prerequisite relations\cite{scheines2014discovering}, aiming to offer a solid foundation of professional knowledge or expertise\cite{backfisch2020professional}. Specifically, we construct an ordered list of concepts for each scholar and only select researchers with more than 50 publications to ensure the sufficient length of their concept lists and the effectiveness of the embedding. Fig. \ref{concept_representation_learning} illustrates this approach for author $A_i$: for any work (paper) $W_t$ published at time $t\in \left[ 1,k \right] $, we list its associated concepts $W _ { t } = \left\{ C _ { 1 } ^ { t } , C _ { 2 } ^ { t } , \cdots C _ { i } ^ { t } , \cdots C _ { n } ^ { t } \right\}$, and these concepts are then chronologically ordered as $A _ { i } = \left\{ W _ { 1 } , W _ { 2 } , \cdots , W _ { k } \right\}$ according to the publication year of the papers. This sequence of concepts effectively encodes the interdependent precedence relationships between them.

We further learn the embedding space based on the established list of trajectories for knowledge navigation.
We applied the Word2Vec embedding model\cite{peng2021neural,mikolov2013efficient} to train vectors for each concept by taking the trajectory of millions of selected authors $\mathrm { C } = \left\{ A _ { 1 }, A _ { 2 }, \cdots, A _ { N } \right\}$ as input, which is shown in Fig. \ref{concept_representation_learning}. Achieving a balance between dimensionality and effectiveness is crucial, especially given our relatively small database of approximately 65,000 concepts. Therefore, we select the smallest dimension possible to avoid sparsity and preserve high semantic information. According to the normalized embedding loss $L(d)$ under different dimensions $d$ proposed in\cite{gu2021principled}, a dimension of 22 is small enough to be efficient and large enough to be effective. To enhance our model's performance, we empirically chose a vector dimension of 24. This slightly exceeds the minimum requirement of 22 suggested in the literature, ensuring both computational efficiency and model effectiveness, particularly suited to the limited size of concepts in OpenAlex.

\subsection{Representation of conceptual navigation space}
To better visualize the structure of the learned embedding space, we use Uniform Manifold Approximation and Projection (UMAP) for dimensionality reduction. The 2D map visualization of concept embeddings through UMAP is shown in Fig. \ref{figure_2}. For clarity and convenience, each discipline is represented by a unique color and assigned a corresponding number on the left side of the figure. This visualization corroborates the effectiveness of our embeddings in capturing the disciplinary classification of concepts. In Fig. \ref{figure_2_19_disciplines}, we colored each concept in $S^+$ according to the color of its discipline root concept. Thus, dots with the same color correspond to concepts classified to the same discipline root, and we marked the discipline number at the point with the highest density in each color to correspond to the respective discipline. We see that there is an apparent disciplinary clustering within the same color, and different clustering exhibits clear disciplinary boundaries. Furthermore, in Fig. \ref{figure_2_multi}, for the ``Multi-interdisciplinary'' concepts in $M^-$ (with multiple indistinguishable root disciplines) are marked with red color, as expected, they are distributed across diverse disciplines without clear clustering modules. 

\begin{figure*}[h]
 \centering
  \begin{minipage}{0.24\textwidth}
     \centering
     \begin{subfigure}{\linewidth}
        \centering
        \includegraphics[width=\linewidth]{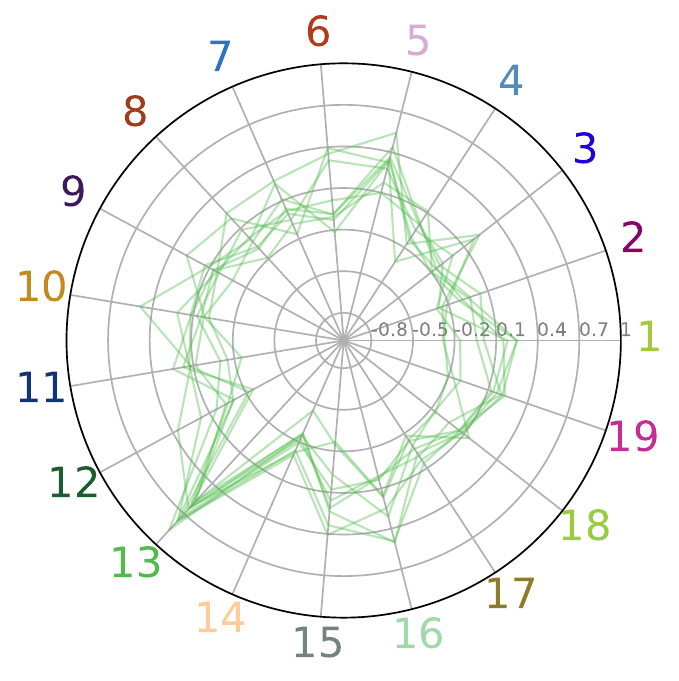}
        \caption{Level 1 Mathematics (13)}
        \label{math_radar_level_1_to_level_0}
     \end{subfigure}
     \vfill
     \begin{subfigure}{\linewidth}
        \centering
        \includegraphics[width=\linewidth]{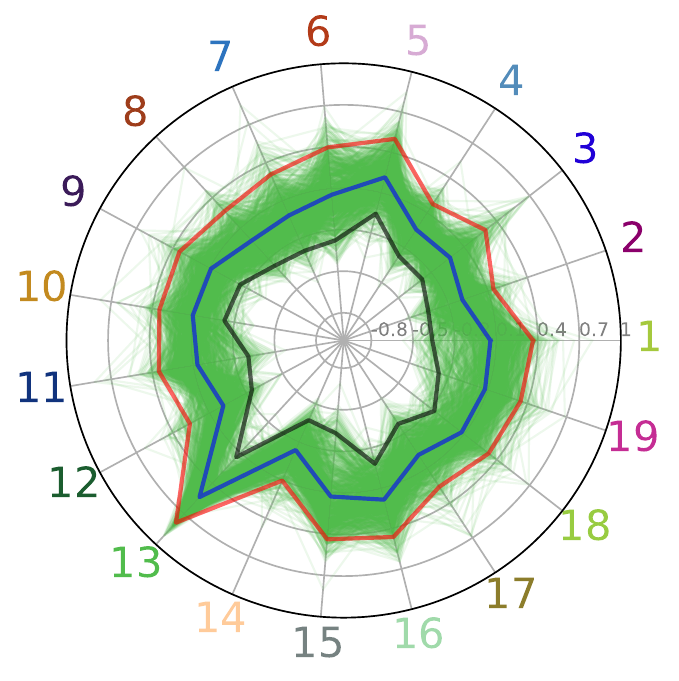}
        \caption{Level 2 Mathematics (13)}
        \label{math_radar_level_2_to_level_0}
     \end{subfigure}
 \end{minipage}
 \hfill
 \begin{minipage}{0.24\textwidth}
     \centering
     \begin{subfigure}{\linewidth}
        \centering
        \includegraphics[width=\linewidth]{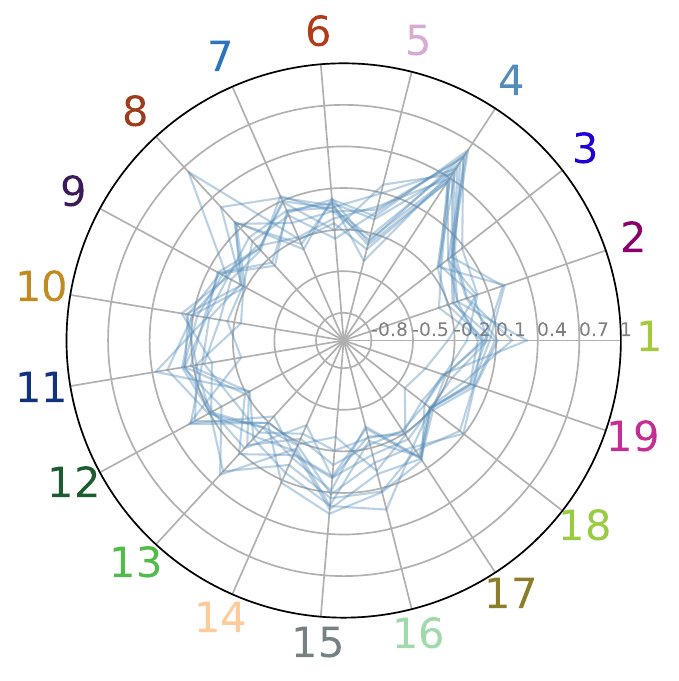}
        \caption{Level 1 Chemistry (4)}
        \label{chemistry_radar_level_1_to_level_0}
     \end{subfigure}
     \vfill
     \begin{subfigure}{\linewidth}
        \centering
        \includegraphics[width=\linewidth]{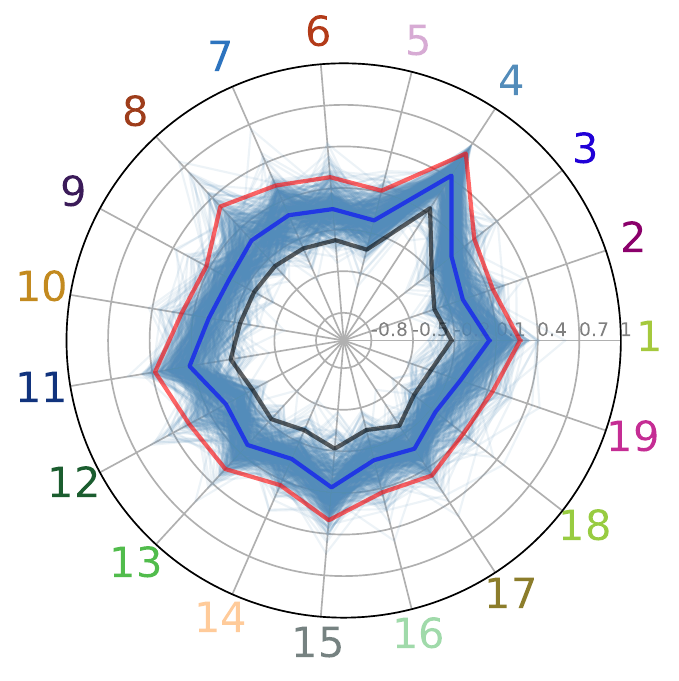}
        \caption{Level 2 Chemistry (4)}
        \label{chemistry_radar_level_2_to_level_0}
     \end{subfigure}
 \end{minipage}
 \hfill
 \begin{minipage}{0.49\textwidth}
     \centering
     \begin{subfigure}{\linewidth}
        \centering
        \includegraphics[width=\linewidth]{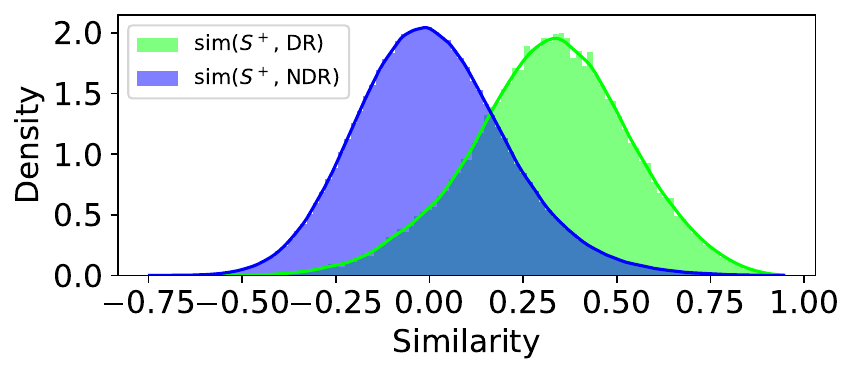}
        \caption{Ancestor propensity validation for concepts in $S^+$}
        \label{sim_with_single_ancestor}
     \end{subfigure}
     \vfill
     \begin{subfigure}{\linewidth}
        \centering
        \includegraphics[width=\linewidth]{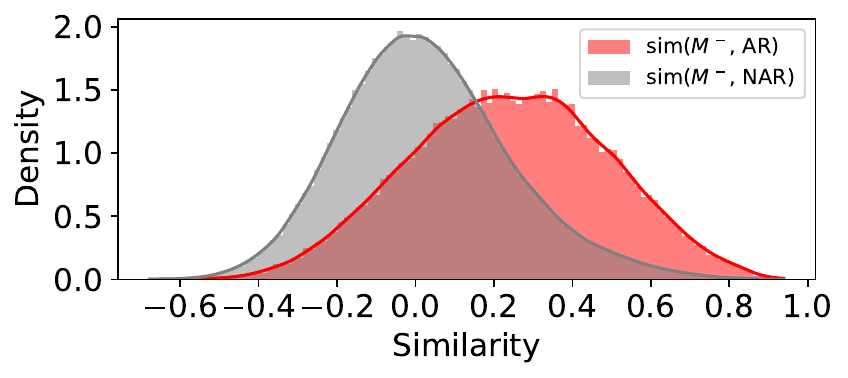}
        \caption{Ancestor propensity validation for concepts in $M^-$}
        \label{sim_with_multiple_ancestors}
     \end{subfigure}
 \end{minipage}
 \caption{\textbf{Propensity validation of concepts with discipline classification}. (a) The vector similarity between level 1 sub-concepts ($c\in S$) of Mathematics and the 19 root concepts. The disciplines are encoded by numbers listed in \ref{figure_2_19_disciplines}, thus values on each axis are the cosine similarities between sub-concepts and corresponding discipline. (b) The vector similarity between level 2 sub-concepts ($c\in S$) of Mathematics and the 19 root concepts. The plot verifies that vectors of all math sub-concepts are semantically closer to discipline Mathematics. The 5-th, 50-th, and 95-th quantiles are marked with black, blue and red lines respectively from center to outside circle. (c) and (d) are another example of discipline propensity verification for level 1 and level 2 sub-concepts of discipline Chemistry. (e) Comparison of two similarity distributions of $S^+$ concepts, $S^+$ concepts and discipline root (DR) $\text{sim}(S^+,\text{DR})$ vs. $S^+$ concepts and its non-discipline root (NDR) $\text{sim}(S^+,\text{NDR})$. (f) Comparison of two similarity distributions of $M^-$ concepts, $M^-$ concepts and its ancestor roots (AR) $\text{sim}(M^-,\text{AR})$ vs. $M^-$ concepts and its non-ancestor roots (NAR) $\text{sim}(M^-,\text{NAR})$. Both plots in (e) and (f) exhibit a significantly higher similarity with their ancestor roots compared with their non-ancestor roots.}
    \label{figure_3}
\end{figure*}

\subsection{Discipline propensity validation of concepts}
To verify the degree of propensity between a given concept $c$ and its classified discipline $d$, we employ the metric of cosine similarity for concepts $i$ and $j$, which calculated as
\begin{equation}
    \text{sim}\left(\vec{i},\vec{j}\right)=\left( \vec{i}\cdot \vec{j} \right) /\left( \mid \vec{i} \mid\times \mid \vec{j} \mid \right),
\end{equation}
where $\vec{i}$ ($\vec{j}$) represents the embedding vector of concept $i$ ($j$), and the metric will be denoted as $s_{ij}$ in subsequent text for simplicity. Specifically, for classifiable concept $c \in S^+$, the single classifiable AR is denoted the discipline root (DR), and the rest 18 roots are denoted as non-discipline roots (NDR). For convenience, we use the number to represent the corresponding discipline. Consider the concept ``PPADS''$\in S^+$ in   Table \ref{discipline_classification_example} as an example, three ancestors are Biology(2), Chemistry(4), Medicine(14), thus AR=$\left\{2,4,14\right\}$ vs. NAR=$\left\{ 1,2,3,\cdots, 19 \right\} \backslash\left\{ 2,4,14 \right\}$, as concept ``PPADS'' is classified to discipline Biology(2), hence DR=$\left\{2\right\}$ vs. NDR=$\left\{ 1,2,3,\cdots, 19 \right\} \backslash\left\{ 2 \right\} $, note that NAR $\subseteq$ NDR.

The results of discipline propensity validation were quantitatively represented on a radar map, each one of the 19 polar axes is marked by a number symbolizing a discipline shown in the left side of Fig. \ref{figure_2_19_disciplines}. To simplify our analysis, we consider concepts in $S$ (single root), concepts in $S^+$ (classifiable) and $M^-$ (indistinguishable) respectively. We further divided the 19 root disciplines into two categories, for each classifiable concept $c\in S^+$: DR vs. NDR, and for each indistinguishable concept $c\in M^-$: AR vs. NAR. The reason for this different division is that concepts in $S^+$ are classifiable, one of the ancestor roots (AR) is classified as a single discipline root (DR), thus NAR $\subseteq$ NDR for concepts $c\in S^+$, hence we use DR vs. NDR for concepts in $S^+$ to avoid the possible lack of disciplines in NAR when using general division AR vs. NAR.
We confirmed that whether for concepts in $S$, or concepts in $S^+$, the cosine similarities between concepts and DR are remarkably higher compared to NDR, and for concepts in $M^-$, the cosine similarities between concepts and AR are significantly higher compared to NAR, as illustrated in Fig. \ref{figure_3}. 

For concepts in $S$, consider the discipline ``Mathematics'' as an example, its affiliated level 1 and level 2 descendant concepts ($c\in S$) were expected to demonstrate significantly higher cosine similarity with Mathematics compared to the other 18 disciplines.
We projected the sub-concepts (level 1 and level 2 concepts) of Mathematics onto each direction marked by a number from 1 to 19 on the radar map shown in Fig. \ref{math_radar_level_1_to_level_0} and Fig. \ref{math_radar_level_2_to_level_0} respectively, each number in the radar map corresponding to the root discipline (shown on the left side of Fig. \ref{figure_2_19_disciplines}). The result coincides with the expectation, which is clearly illustrated by the most outward-pointing angle (MOPA) towards the direction of the discipline of Mathematics (13) on the radar map. Both level 1 concepts in Fig. \ref{math_radar_level_1_to_level_0} and level 2 concepts in Fig. \ref{math_radar_level_2_to_level_0} exhibit an obvious MOPA towards its discipline Mathematics, indicating a strong discipline propensity of descendant concepts to its labeled discipline Mathematics (DR) compared with the rest non-discipline roots (NDR). We also show another example of discipline propensity validation for sub-concepts (level 1 and level 2) of discipline Chemistry respectively shown in Fig. \ref{chemistry_radar_level_1_to_level_0} and Fig. \ref{chemistry_radar_level_2_to_level_0}, both radar maps exhibit an obvious MOAP towards discipline Chemistry (DR) compared with non-discipline roots (NDR), also supporting the conclusion. Examples of radar mappings for propensity validation of level 1 concepts across various disciplines are shown in Fig. \ref{radar_mappings_all_disciplines_level_1}. Due to the limited number of sub-concepts in Environmental Science, the radar mapping for this discipline has been excluded. Additionally, radar mappings for level 2 concepts in other disciplines are presented in Fig. \ref{radar_mappings_all_disciplines_level_2}.

\begin{figure*}[htbp]
    \centering
    \begin{subfigure}[b]{\textwidth}
        \centering
        \includegraphics[width=\textwidth]{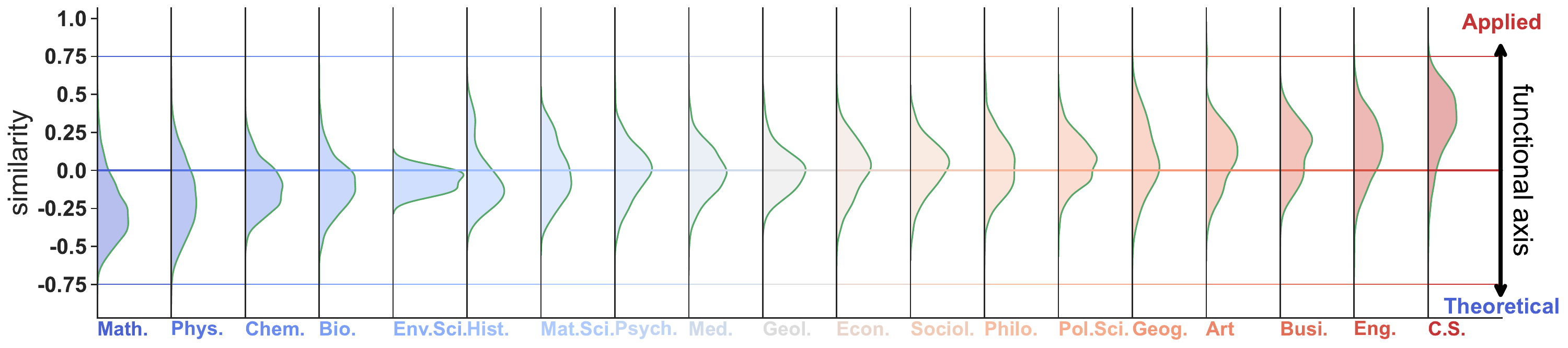}
        \caption{Functional axis between Theoretical and Applied}
        \label{functional_axis_theoretical_applied}
    \end{subfigure}
    \begin{subfigure}[b]{\textwidth}
        \centering
        \includegraphics[width=\textwidth]{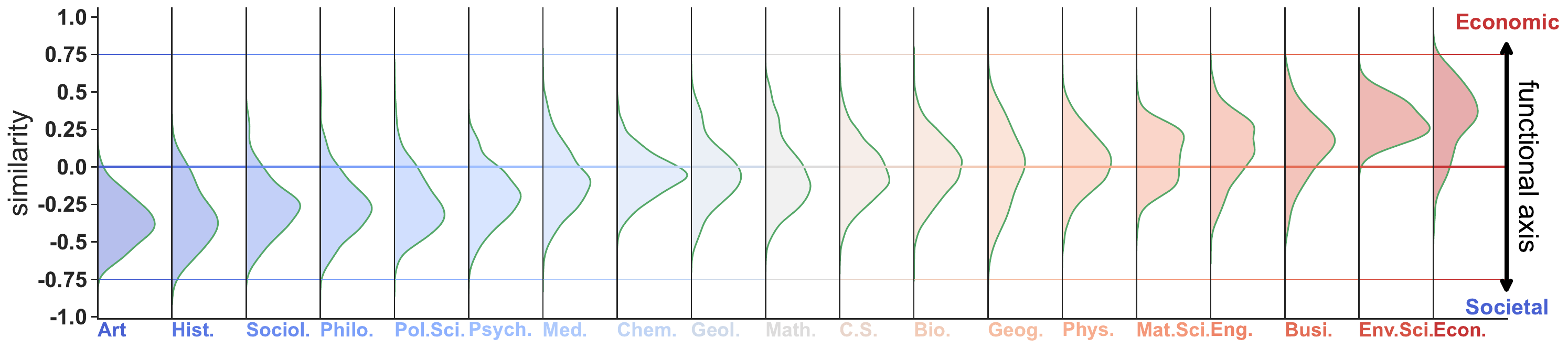}
        \caption{Functional axis between Societal and Economic}
        \label{functional_axis_societal_economic}
    \end{subfigure}
    \caption{\textbf{Functionality projections of knowledge from 19 disciplines on predefined axes}. To avoid overlap caused by lengthy terms, we use abbreviations for discipline categories listed in   Table \ref{discipline_abbreviations} in SI. (a) Projection of all disciplines on functional axis from Theoretical (Mathematics and Physics) to Applied (Engineering and Computer science). (b) Projection of all disciplines on functional axis from Societal to Economic.}
    \label{functionality_projections}
\end{figure*}

To further verify the conclusion for concepts from all classifiable concepts in $S^+$ in stead of concepts with a single discipline (Fig. \ref{math_radar_level_1_to_level_0}$\sim$Fig. \ref{chemistry_radar_level_2_to_level_0}). 
We present the distribution of cosine similarities between $S^+$ concepts and its discipline roots (DR) labeled as $\text{sim}(S^+,\text{DR})$, compared with the distribution of cosine similarities between $S^+$ concepts and its non-discipline roots (NDR) labeled as $\text{sim}(S^+,\text{NDR})$.
The noticeable rightward shift of the $\text{sim}(S^+,\text{DR})$ distribution (green) compared to the $\text{sim}(S^+,\text{NDR})$ distribution (blue) indicates a significantly higher similarity of $S^+$ concepts with its discipline roots (DR).

In comparison with classifiable concepts in $S^+$, we further validate the similarity distributions of indistinguishable concepts in $M^-$ shown in Fig. \ref{sim_with_multiple_ancestors}. The distribution of cosine similarities between $M^-$ concepts and its ancestor roots (AR) labeled as $\text{sim}(M^-,\text{AR})$, compared with the distribution of cosine similarities between $M^-$ concepts and its non-ancestor roots (NAR) labeled as $\text{sim}(M^-,\text{NAR})$. The result indicates that, even for indistinguishable concepts in $M^-$, a similar pattern is observed for $M^-$ concepts, where the $\text{sim}(M^-,\text{AR})$ distribution with red color, exhibits a significant right shift pattern compared with the $\text{sim}(M^-,\text{NAR})$ distribution with gray color. These distributions confirm that both $S^+$ concepts and $M^-$ concepts, as well as $S$ concepts exhibit a significantly higher similarity with their ancestor disciplines compared to non-ancestors roots.

\subsection{Functionality projections of knowledge}
To further explore the distinctive functionality variance among 19 disciplines across a predefined conceptual dimension, we calculate the cosine similarity between concept embeddings and a specially designed conceptual axis. By analyzing the positioning of concepts from each discipline along this functional dimension, we observe trends of distributional shifts and acquire insights into their diverse functional properties. The distributions of various disciplines highlight their characteristics and unique functional roles, revealing patterns in how the semantic functionality of concepts converges within and diverges across disciplines.

To construct the conceptual axis that distinguishes different functionality attributes, we categorize 19 disciplines into eight major functional groups: Theoretical, Applied, Chemical, Biomedical, Societal, Economic, Humanities, and Geographical. These groups represent distinct functionalities, as detailed in Table \ref{discipline_groups} in SI. Consider the functional groups Theoretical and Applied as an example. The functional group of Theoretical encompasses concepts classified as Mathematics and Physics, demonstrating the property of the theoretical attribute. In contrast, the Applied group covers concepts in Computer Science and Engineering, indicating the application attributes of these disciplines. The other six functional groups are Chemical, Biomedical, Societal, Economic, Humanities, and Geographical. The detailed disciplines included in these functional groups are shown in   Table \ref{discipline_groups}.
Secondly, we calculated the average of the embedding vectors of two functional groups to construct the functional axis. Let $G$ be the set of all concepts a functional group, the average embedding of group $G$ is represented as $\vec{G}=\frac{1}{\left| G \right|}\sum_{c\in G}{\vec{c}}$, and the conceptual dimension from group $G_1$ to group $G_2$ is $\overrightarrow{G_2}-\overrightarrow{G_1}$. We then established functional axes between these concept groups, such as the axis from Theoretical to Applied, or the axis from Societal to Economic.

\begin{figure*}[h]
    \centering
    \begin{minipage}{0.328\textwidth}
         \centering
         \begin{subfigure}{\linewidth}
     \centering
     \includegraphics[width=\linewidth]{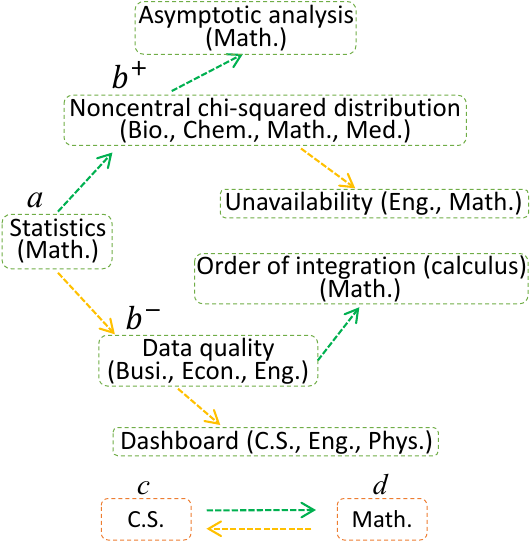}
     \caption{Analogy inference 1}
     \label{analogy_inference_1}
         \end{subfigure}
    \end{minipage}
     \begin{minipage}{0.328\textwidth}
         \centering
         \begin{subfigure}{\linewidth}
     \centering
     \includegraphics[width=\linewidth]{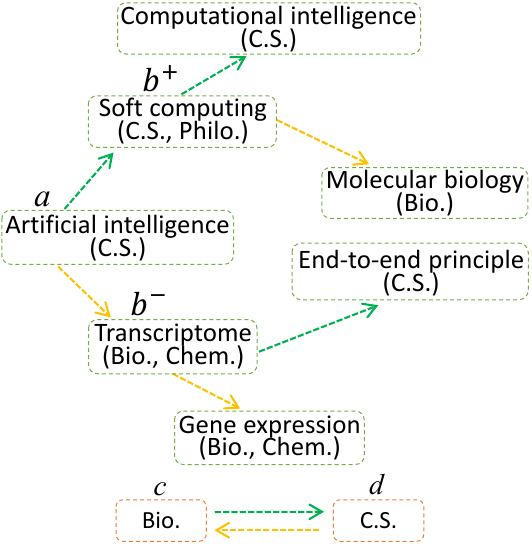}
     \caption{Analogy inference 2}
     \label{analogy_inference_2}
         \end{subfigure}
     \end{minipage}
    \begin{minipage}{0.328\textwidth}
        \centering
        \begin{subfigure}{\linewidth}
     \centering
     \includegraphics[width=\linewidth]{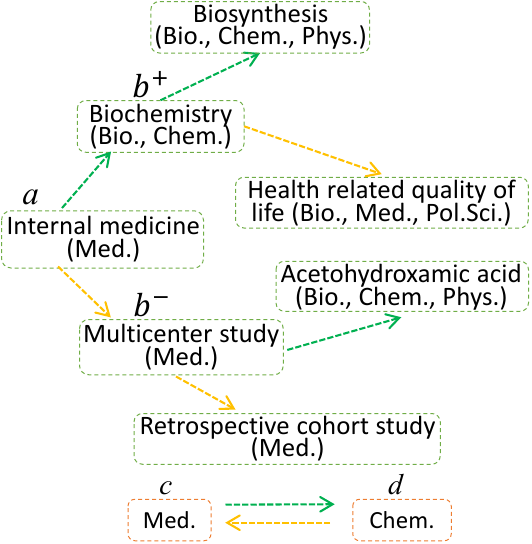}
     \caption{Analogy inference 3}
     \label{analogy_inference_3}
        \end{subfigure}
    \end{minipage}
    \caption{\textbf{Examples of analogy inference}. (a) Analogy inference example with inference axis between Mathematics and Computer science while the seed concept is Statistics. (b) Analogy inference example with inference axis between Computer science and Biology while the seed concept is Artificial intelligence. (c) Analogy inference example with inference axis between Chemistry and Medicine while the seed concept is Internal medicine.}
    \label{analogy_inferences}
\end{figure*}

The functional axis allows us to project concepts from each discipline onto this axis. For instance, consider the axis from Theoretical to Applied and the axis from Societal to Economic respectively, we evaluated the alignment of various disciplines with this axis by plotting the cosine similarity distribution of their concepts against this axis, the results are shown in Fig. \ref{functionality_projections}. Fig. \ref{functional_axis_theoretical_applied} shows the positioning of PDFs for all disciplines along the axis from Theoretical to Applied, with Mathematics concepts positioned at the more Theoretical end and Computer Science at the Applied end. On the contrary, concepts from other disciplines occupy different positions in this dimension based on their theoretical application level, confirming the effectiveness of this functional axis in differentiating disciplines. 

Similarly, in the axis from Societal to Economic, as depicted in Fig. \ref{functional_axis_societal_economic}, demonstrates the range of concepts from Societal fields like Sociology, Political science and Psychology to Economic fields like Economics and Business. It highlights how disciplines such as Art, History align more with Societal sciences, while fields like Engineering and Environmental science are closer to Economic sciences. We also displayed two other examples of functional projections, as shown in Fig. \ref{functionality_projections_SI}. These examples respectively illustrate the functional distributions between Chemical and Biomedical fields, as well as Humanities and Geographical fields.
Collectively, these axes offer a comprehensive view of the functional trends and relationships among concepts across a spectrum of academic disciplines.

Overall, we examined and verified the embedding representations of knowledge from various disciplines and analyzed the semantic functionality along the crafted conceptual dimensions. This provides an intuitive understanding of the similarities and differences in domain knowledge across various fields. In the next section, we utilize the semantics of knowledge to demonstrate meaningful applications, such as analogy inference and navigation.

\section{Results}\label{result_section}
In this section, we propose the Scientific Concept Navigator (SciConNav), and demonstrate its impactful applications in analogy inference and knowledge navigation. SciConNav addresses challenges in selecting suitable research topics and inferring interdependent learning pathways. Firstly, we employed multi-step analogy inference to explore analogical concepts along predefined axes, resulting in meaningful inference graphs. Secondly, we conducted global knowledge navigation to examine the accessibility between knowledge domains. Lastly, we highlighted the critical role of interdisciplinary concepts in the global knowledge network.

\subsection{Analogies of interconnected knowledge}
The SciConNav model facilitates analogical reasoning between scientific concepts, which infer interconnected knowledge in a predetermined manner through analogy, helping researchers identify highly relevant topics based on their current research background. In this context, analogy is understood through vector relationships in the embedding space, guided by the principle ``$a$ is to $b$ as $c$ is to $d$''\cite{drozd2016word}. A well-known illustration of this principle is the analogy ``king is to ? as man is to woman''\cite{ethayarajh2018towards}, which can be solved using vector arithmetic for analogical reasoning. Concept analogy inference is to infer related concepts with possible functional dependency or causal relationship from seed concepts along positive and negative analogy dimensions. Here, we treat concepts as analogy objects; let $\vec{a}$ represents the embedding vector of concept $a$, we defined $c$ is to $d$ as the direction vector $\overrightarrow{cd}=\vec{d} - \vec{c}$, hence the inferred concepts  
\begin{equation}
    b^{\pm}=arg\underset{b}{\min}\left\{ \text{sim}\left( \vec{b},\vec{a}\pm \overrightarrow{cd} \right) \right\}, 
\end{equation}
can be obtained via vector arithmetic along the positive direction ($b^+$) and negative direction ($b^-$).

\begin{figure*}[ht]
    \centering
    
    \begin{subfigure}[c]{0.46\textwidth}
        \centering
        \includegraphics[width=\linewidth]{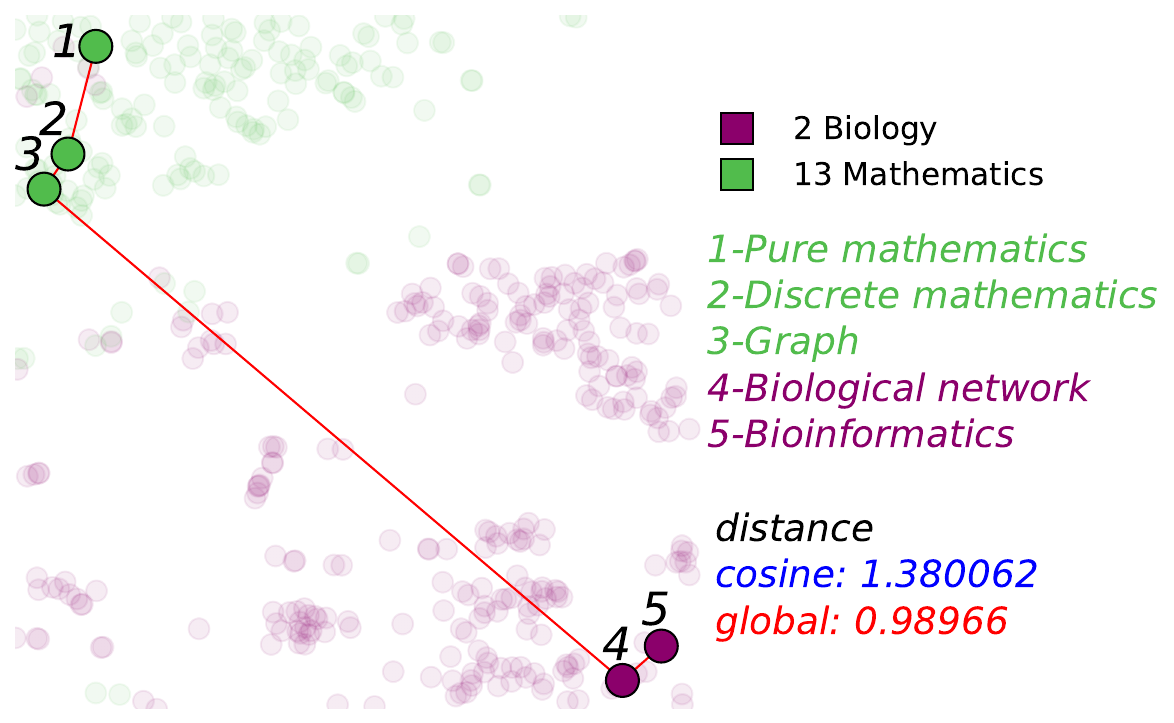}
        \caption{Global navigation between Pure Math. and Bioinformatics}
        \label{path_Bioinformatics_Pure_mathematics}
    \end{subfigure}
    \begin{subfigure}[c]{0.45\textwidth}
        \centering
        \includegraphics[width=\linewidth]{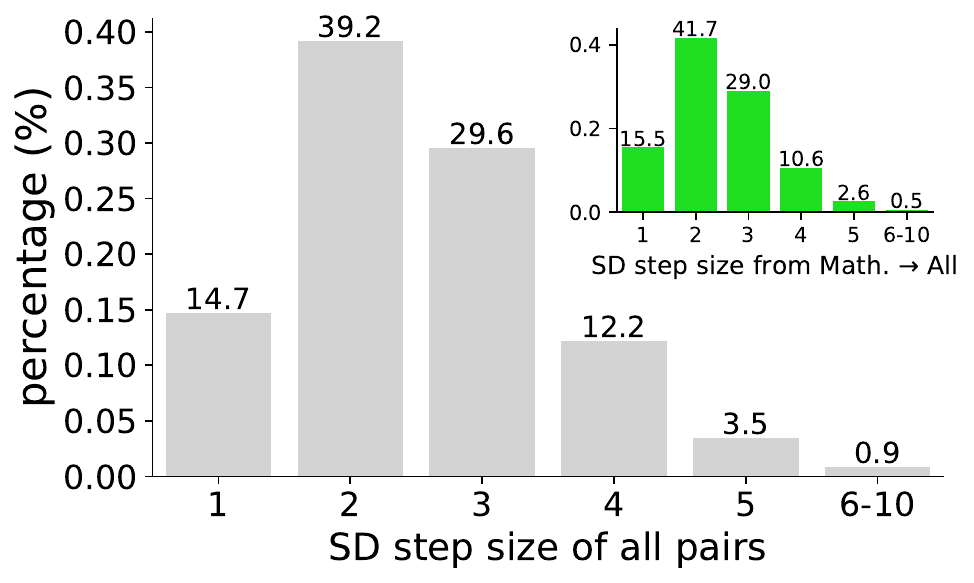}
        \caption{Step size distribution}
        \label{step_size_distribution}
    \end{subfigure}
    \begin{subfigure}[t]{0.8\textwidth}
        \centering
        \includegraphics[width=\linewidth]{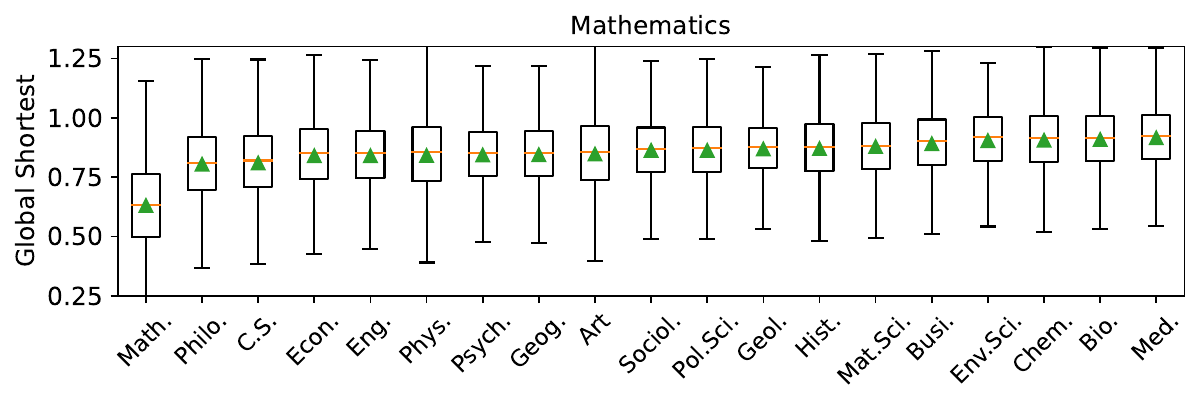}
        \caption{Knowledge accessibility of concepts form Mathematics to 19 disciplines}
        \label{mathematics_shortest_distances}
    \end{subfigure}
    \caption{\textbf{Global knowledge navigation and network accessibility}. (a) Illustrates the shortest path from ``Pure mathematics'' to ``Bioinformatics'', highlighting the cross-disciplinary connections. (b) Global step size distribution from concepts in Mathematics to concepts in all 19 disciplines (green) and the distribution for all pairwise concepts (lightgray). (c) Histogram of step sizes for the shortest cosine distance paths, with 99\% under 5 steps.}
    \label{knowledge_navigation_and_accessibility}
\end{figure*}

We show a two-step concept analogy inference in Fig. \ref{analogy_inference_1}, where the analogy axis lies between Computer science and Mathematics. From the seed concept ``Statistics,'' with the positive direction, we reach ``Noncentral chi-squared distribution,'' and then starting from ``Noncentral chi-squared distribution,'' we reach ``Asymptotic analysis'' and ``Unavailability'' in the opposite direction, from ``Statistics'' with the negative direction we reach ``Data quality,'' and then starting from ``Data quality,'' we reach ``Order of integration (calculus)'' in the positive direction and ``Dashboard'' in the opposite. In Fig. \ref{analogy_inference_2}, the inference axis lies between ``Computer science'' and ``Biology'' while the seed is ``Artificial intelligence'', and in Fig. \ref{analogy_inference_3}, the inference axis lies between ``Chemistry'' and ``Medicine'' while the seed concept is ``Internal medicine''.

From the examples above, we demonstrate that conceptual analogical reasoning enhances the understanding and reasoning capabilities of NLP systems by identifying and utilizing correlations, hierarchies, and causal relationships between knowledge concepts. Each inferred new concept can be considered a likely choice for unexplored topics in cross-disciplinary research or learning. This, in turn, improves the intelligence and accuracy of applications such as information retrieval, education, and question-answering systems. Additional examples of two-step analogy inference are presented in Fig. \ref{more_analogy_inference}.

\subsection{Global knowledge navigation and network accessibility}
The concepts of analogy inference and functionality inspired us to investigate global connections across all disciplines through the shortest path, extending beyond the typical focus on local similarity through analogy inference. This approach addresses the challenge of identifying interdependent learning pathways between selected topics. SciConNav constructs a fully connected network using the top $n$ (=20,000) representative concepts with the highest number of related papers, ensuring that each selected concept has at least 3,410 associated works. The cosine distance $w_{ij}=1-s_{ij}$ between concept $i$ and concept $j$ serves as the edge weight. For each pair of concepts, we calculate the pairwise weighted shortest path distance, record the path step length for each weighted shortest path, and further analyze the shortest distance (SD) distribution from Mathematics to all other disciplines. The results of global knowledge navigation are shown in Fig. \ref{knowledge_navigation_and_accessibility}.

In Fig. \ref{path_Bioinformatics_Pure_mathematics}, we present the example of knowledge navigation with the shortest cosine distance path, which traces a route between ``Pure mathematics'' (a concept in Mathematics) and ``Bioinformatics'' (a concept in Biology) bi-directionally. The global SD along this path is calculated to be 0.99, which is significantly shorter than the direct cosine distance between these two concepts, calculated at 1.38. This notable reduction in distance is achieved by connecting multiple globally relevant concepts along the path. These intermediate concepts serve as bridges connecting ``Pure mathematics'' and ``Bioinformatics''. They not only shorten the distance between these two distant concepts but also highlight the importance of knowledge connections in facilitating academic exploration and discovery. Moreover, this shortest path reflects a degree of prerequisite correlation, suggesting that concepts lying on these paths encapsulate temporal relationships within the historical sequences of knowledge development. Therefore, the shortest path method does more than just connect different concepts; it uncovers meaningful and interpretable global connections that span across diverse academic disciplines.

Denote $d_{ij}$ as the SD between two concept $i$ and $j$, we calculated the SD route step size of all pair concepts from 19 disciplines, and obtained the step size histogram in Fig. \ref{step_size_distribution} with gray color. We showed that for all the pairwise shortest path steps, we observed that 99\% of the step sizes in the SD paths across all areas are less than 5, indicating a high level of interconnectedness within a relatively small number of steps. For concepts in Mathematics, we further showcase SD step size histogram between sub-concepts from Mathematics and concepts from 19 disciplines, most of these SD path lengths are short in most cases, encompass 2 or 3 steps, shown in inset of Fig. \ref{step_size_distribution} with green color, which means they only need 1 or 2 intermediate concepts to connect the source and the target concepts.

Denote $d_{ij}$ as the shortest distance (SD) between two concepts $i$ and $j$. We calculated the SD route step size for all pairs of concepts across 19 disciplines and obtained the step size histogram shown in Fig. \ref{step_size_distribution} in gray. Our analysis revealed that 99\% of the step sizes in the SD paths across all areas are less than 5, indicating a high level of interconnectedness within a relatively small number of steps. Specifically for concepts in Mathematics, we further examined the SD step size histogram between sub-concepts in Mathematics and concepts from the 19 disciplines. Most of these SD path lengths are short, typically encompassing 2 or 3 steps, as shown in the inset of Fig. \ref{step_size_distribution} in green. This indicates that only 1 or 2 intermediate concepts are needed to connect the source and target concepts.

Our analysis reveals that sub-concepts within the discipline of Mathematics display the shortest distances to sub-concepts in the Mathematics domain itself. When examining the relationships between Mathematics and the other 18 disciplines studied, we found no significant differences in distances. However, among these other disciplines, Philosophy demonstrates the second shortest distance to Mathematics, followed closely by Computer Science. This suggests that concepts in Mathematics are more closely connected to the field of Philosophy and Computer Science in scientific research. These findings highlight the interdisciplinary nature of mathematical concepts and their particular relevance to philosophical inquiry and computational sciences. 

By mapping these conceptual pathways, we provide a structured approach to navigating the intricate web of scientific knowledge. This framework offers a logical, step-by-step learning trajectory across diverse fields. Our focus on global shortest paths between concepts addresses key questions: (1) Which concepts serve as critical bridges between distinct disciplines? (2) How can researchers efficiently transition from their current focus to explore new topics incrementally? These insights are crucial for understanding the interconnected nature of knowledge progression, guiding researchers in strategic research planning, and facilitating interdisciplinary collaboration and innovation. Our approach not only reveals the most efficient routes for knowledge acquisition but also highlights potential areas for cross-disciplinary breakthroughs, researchers can more effectively expand their expertise, identify novel research directions, and contribute to the advancement of science across traditional boundaries.

\begin{figure*}[h]
    \centering
    \begin{subfigure}[m]{0.48\textwidth}
        \centering
        \includegraphics[width=\linewidth]{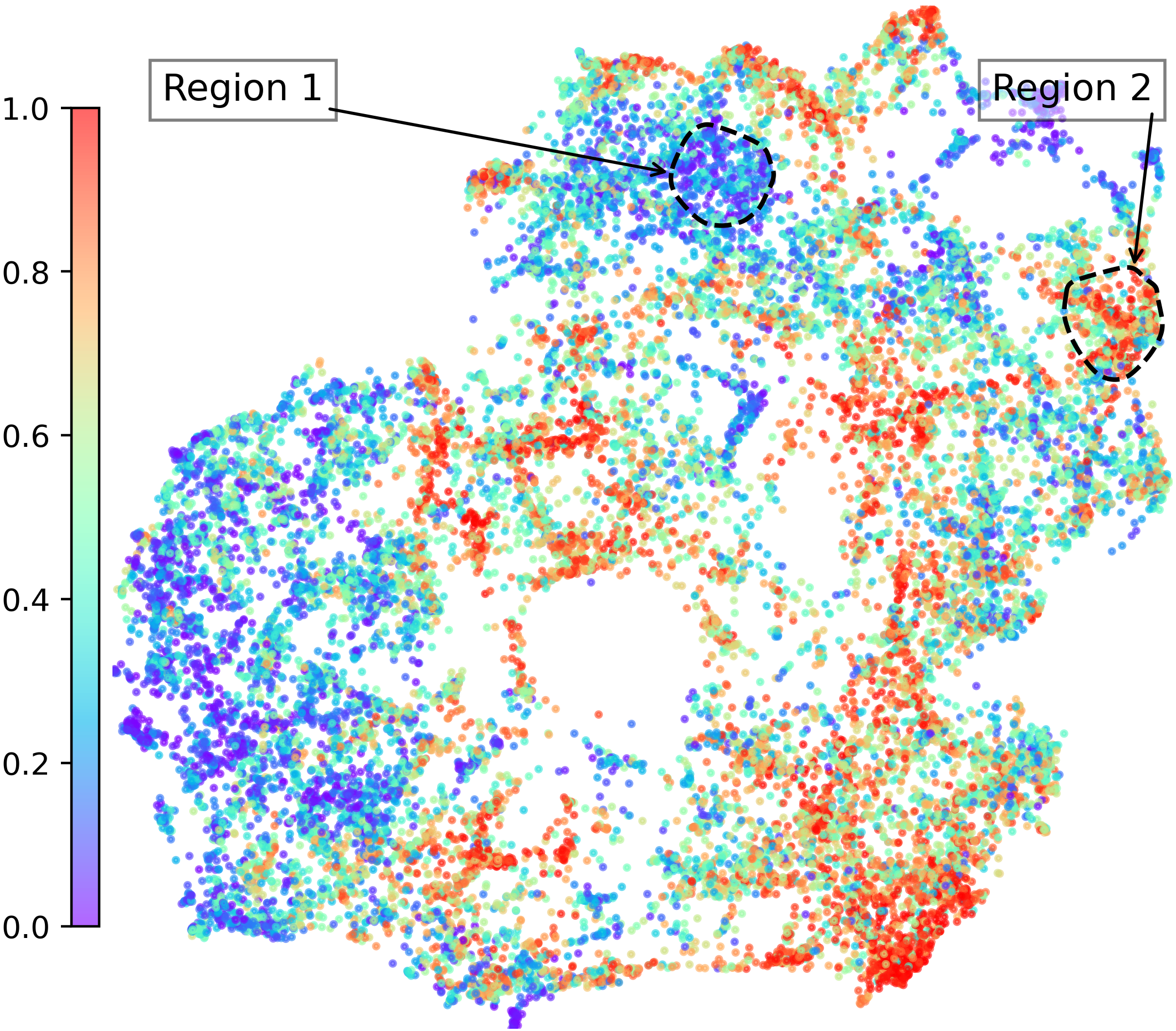}
        \caption{Heatmap of concept closeness centrality}
        \label{closeness_centrality}
    \end{subfigure}
    \begin{subfigure}[m]{0.45\textwidth}
        \centering
        \begin{subfigure}{\textwidth}
     \centering
     \includegraphics[width=\linewidth]{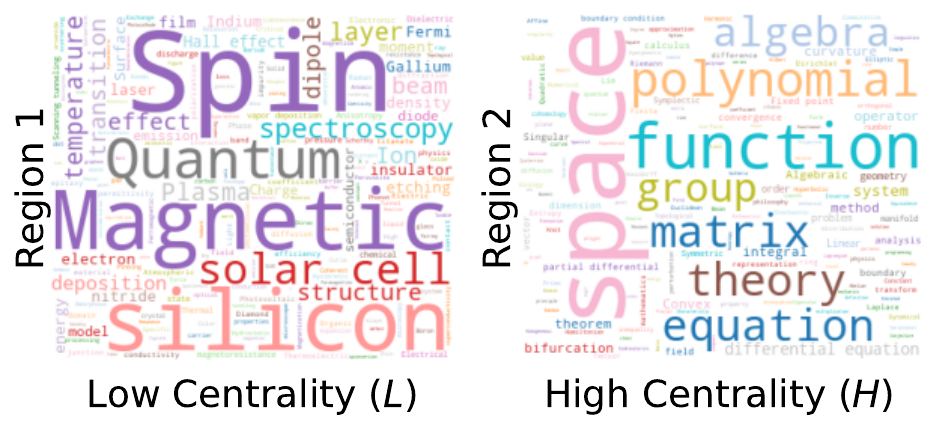}
     \caption{Concept word clouds of two selected small regions}
     \label{closeness_word_cloud}
        \end{subfigure}
        \begin{subfigure}{\textwidth}
     \centering
     \includegraphics[width=\linewidth]{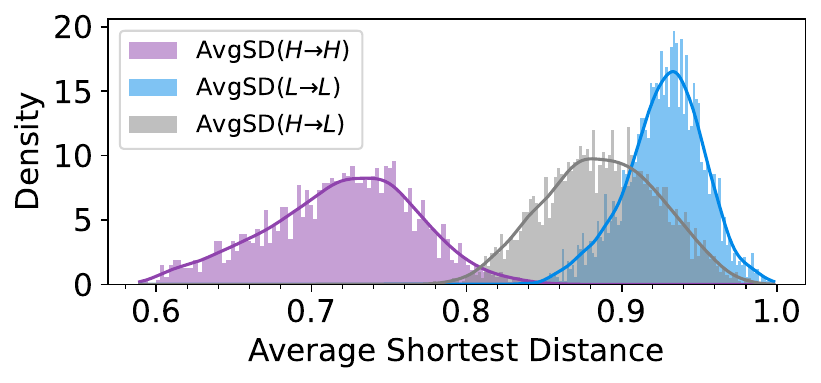}
     \caption{AvgSD between 2k $H$ and 2k $L$ concepts by closeness}
     \label{ave_closeness_SD_pdf}
        \end{subfigure}
    \end{subfigure}
    \caption{\textbf{Centrality measure the importance of interdisciplinary concepts}. (a) Embedding map of 20,000 key concepts, color-coded by Closeness centrality. (b) Concept word clouds of two selected regions, the Region 1 mainly contains concepts with low closeness centrality, Region 2 mainly contains concepts with high closeness centrality.}
    \label{closeness_centralities}
\end{figure*}

\subsection{The bridging role of interdisciplinary concept}
The centrality, or the accessibility of a concept indicates its significance. In our global knowledge network among selected 20,000 concepts, we focus on two types of centralities: closeness and betweenness. Closeness centrality is calculated as the inverse of the average SD from the focal to other concepts, measures the proximity of a concept to all others, indicating its efficiency in interacting with a wide range of concepts across different fields. Betweenness centrality is calculated as the fraction of the shortest paths that pass through the focal concept, measures a concept's role in information transfer, acting as bridges between different clusters or disciplines. We denote $H$ as the set of selected concepts with high centrality and $L$ as the set of selected low centrality concepts. We further visually highlighted concepts with high closeness centrality values in red on our embedding maps in Fig. \ref{closeness_centralities}. 

\begin{table}[h]
    \centering
    \caption{Interdisciplinary odds ratio of top centrality concepts.}
    \begin{tabular}{c|ccc|ccc}
        \hline & \multicolumn{3}{|c|}{ Closeness centrality } & \multicolumn{3}{c}{ Betweenness centrality } \\
        \hline Top & $\in M$ & $\in S$ & Odds & $\in M$ & $\in S$ & Odds \\ \hline 
        200 & 113 & 87 & 1.23 & 113 & 87 & 1.30 \\
        500 & 291 & 209 & 1.39 & 315 & 185 & 1.70 \\
        1000 & 644 & 356 & 1.81 & 667 & 333 & 2.00 \\
        1500 & 987 & 513 & 1.92 & 1030 & 470 & 2.19 \\
        2000 & 1333 & 667 & 2.00 & 1398 & 602 & 2.32 \\
        \hline
    \end{tabular}
    \label{interdisciplinary_odds_ratio}
\end{table}

Our analysis revealed a distinct pattern that concepts with high closeness centrality tend to cluster within specific fields shown in Fig. \ref{closeness_centrality}, in contrast to the concepts with high betweenness centrality, which are more widely distributed across various disciplines (as illustrated in Fig. \ref{betweenness_centralities} in SI). We further found that interdisciplinary concepts ($M$) with a considerable proportion play important roles in the knowledge network. Among the top 1\% concepts with the highest closeness centrality values, we identified 200 concepts (top 200 $H$ concepts), of which 113 are interdisciplinary $M$ concepts shown in  Table \ref{interdisciplinary_odds_ratio} (same for betweenness centrality). The significance is illustrated by the increasing proportion of interdisciplinary $M$ concepts compared to $S$ concepts as the number of top $H$ concepts increases, as indicated by the odds ratios shown in   Table \ref{interdisciplinary_odds_ratio}. Both the closeness and betweenness centrality measures showed a large fraction of interdisciplinary concepts with high odds ratios among the top-ranked concepts.

To elucidate the varying centrality patterns, we selected two representative regions with contrasting centralities, showcasing the concentration of main concepts through word clouds in Fig. \ref{closeness_word_cloud}. Our findings indicate that $L$ concepts in Region 1 with low accessibility, primarily from natural sciences and technological domains, are specialized within their fields. These include fundamental physics (spin, quantum phenomena, magnetic properties), materials science (silicon), energy technology (solar applications), and biological fundamentals (cellular structures). In contrast, $H$ concepts in Region 2, primarily fundamental mathematical and theoretical constructs, serve a bridging role, aligning with the foundational role of basic sciences like mathematics in real-world research. This includes spatial concepts, functional analysis, algebraic structures (polynomials, groups), matrix theory, and advanced methods like differential equations. These cornerstone $H$ concepts exhibit high centrality due to their widespread applications and connections across various scientific disciplines, crucial for modeling complex systems, underpinning numerous scientific theories, and addressing practical challenges. This juxtaposition of selected $H$ and $L$ regions offers a nuanced understanding of concept importance distribution, revealing how centrality measures reflect the organization of knowledge within the broader scientific landscape and interconnectedness across scientific domains. 

To analyze how tightly connected these concepts are within or across $H$ and $L$ domains, we measure the average internal and cross-domain distance. We selected the top 2,000 $H$ concepts and the bottom 2,000 $L$ concepts from the whole network, including all 20,000 representative concepts. We then calculated the SD for all pair-wise concepts in $H$, denoted as $\left\{d_{ij} \mid i \neq j, i \in H, j \in H\right\}$, and represented the distance distribution as $\text{SD}(H, H)$. Similarly, $\text{SD}(L, L)$ represents the set of shortest distances for all pair-wise concepts in $L$. The results, shown in Fig. \ref{SD_pdf_of_centrality} in SI, indicate that concepts within $H$ have shorter shortest distances compared to those within $L$. Fig. \ref{closeness_SD_pdf} and Fig. \ref{betweenness_SD_pdf} further illustrate the comparisons for closeness and betweenness, respectively. We further define the average shortest distance (AvgSD) of concept $i$ as
\begin{equation}
    \text{AvgSD}\left(i\right) = \frac{1}{\left| R\left( i \right) \right|}\sum_{j\in R\left( i \right)}{d_{ij}}
\end{equation}
where $R(i)$ is the set of reachable concepts from concept $i$. We calculated the AvgSD for each concept $i \in H $ (reachable nodes $R(i)=H$), and represented the distribution of these average shortest distances as AvgSD( $H \rightarrow H )$. Similarly, AvgSD( $L \rightarrow L )$ represents the distribution of average shortest distances from $L$ concepts to themselves (reachable nodes $R(i) = L$). Additionally, we the AvgSD( $H \rightarrow L )$ represents the distribution of AvgSD from $H$ concepts to $L$ concepts (reachable nodes $R ( i ) = L$ for $\forall i \in H $). The results of interconnectivity analysis selected from the whole region with closeness centrality are shown in Fig. \ref{ave_closeness_SD_pdf}, aligns with results from two selected smaller regions shown through word clouds in Fig. \ref{closeness_word_cloud}. We observe that AvgSD$( H \rightarrow H )$ is positioned at the left part, indicating that the AvgSD among high centrality $H$ concepts is shorter, indicating a dense interconnectivity, particularly in fundamental mathematical and theoretical constructs. On the other hand, AvgSD$( L \rightarrow L )$ is placed at the right end, suggesting that low centrality $L$ concepts exhibit larger AvgSD, reflecting less interconnectivity with more specialized scope within domains, such as natural sciences and associated technologies. Finally, AvgSD( $H \rightarrow L )$ is situated in the middle, indicating that the shortest distances from $H$ to $L$ fall between AvgSD$(H \rightarrow H)$ and AvgSD$( L \rightarrow L )$, which verified the conclusion that high closeness $H$ concepts serve as bridges, enhancing interdisciplinary linkage and knowledge integration. 

Our findings reveal that concepts with high closeness centrality tend to cluster within specific fields like Mathematics, while those with high betweenness centrality are more widely distributed across disciplines. 
This interconnectivity is further strengthened by the high proportion of $M$ concepts within the $H$ group. These $M$ concepts not only enrich individual disciplines but also catalyze the emergence of new interdisciplinary fields, demonstrating the dynamic and evolving nature of knowledge. Our analysis shows that interdisciplinary concepts $( M )$ play a pivotal bridging role in the structure and flow of information within the global knowledge network. This is supported by the high proportion of $M$ concepts among the top centrality $( H )$ concepts, as indicated by increasing odds ratios of $M$ concepts over $S$ concepts, highlighting their central position in linking different areas of study. While $H$ concepts are key to the network's structural integrity and bridging role, interdisciplinary $M$ concepts are essential for promoting cross-disciplinary knowledge integration, fostering innovative approaches, and enabling researchers to draw on a broader range of insights and methodologies. Together, they contribute to a robust and dynamic knowledge network. 

\section{Discussion}
In this study, we introduced a novel approach that combines the knowledge of research trajectories with Word2Vec model to learn concept representations. Building on this foundation, we proposed the SciConNav model for concept inference, utilizing the conceptual navigation space of knowledge and network methods. Our approach involved analyzing the research trajectories of millions of scholars. We employed the semantics of concept embeddings to infer the dependency connections across 19 disciplines, enabling us to track the evolution of scientific knowledge and emerging technologies. This provides valuable resources for early-career scientists and crucial insights for knowledge integration and cross-domain retrieval. This approach allowed us to identify key concepts with potential prerequisite relationships, offering a structured and authoritative knowledge base. The dependency connections among concepts, learned from extensive research trajectories, serve as a navigation tool for global knowledge. This foundation accelerates scientific discovery and enhances knowledge navigation, facilitating interdisciplinary collaboration, knowledge integration, and generation.

The SciConNav model with conceptual knowledge space holds significant values in several aspects. First, it provides scientists with a comprehensive overview of scientific knowledge. By analyzing functionality projections along predefined axes, we offer insights into the theoretical and applied aspects of concepts, highlighting relationships and functional differences across disciplines. Additionally, the concept analogy inference, undertaken through a step-by-step local approach, allows for the exploration of cross-disciplinary knowledge and the drawing of analogical relations from diverse fields. This enhances logical reasoning, supports decision-making\cite{schulz2023knowledge}, and fosters creative thinking\cite{wegerif2010exploring,xiong2022influence}. Moreover, the global connections through the shortest cosine distance paths navigate the research progression, offering a customized learning pathway from the current knowledge base to desired knowledge, particularly for researchers less familiar with certain areas. By investigating global accessibility of concepts using centrality measures, we can apprehend the connectivity across disciplines and identify key concepts that highlight the bridging role of interdisciplinary concepts.

Despite these contributions, our study has limitations. The dataset, comprising approximately 65,000 concepts, which are validated by Wiki, could be expanded for greater diversity and depth. Advanced language models could offer richer semantic encoding, and more sophisticated network methods could enhance cross-domain knowledge inference. Additionally, the high dimensionality of the embedding space presents challenges in interpreting the knowledge structure, suggesting a need for more nuanced analytical tools to unpack the black box.

In conclusion, while there is room for improvement and further exploration, our conceptual SciConNav model represents a significant advancement in knowledge navigation and scientific discovery, which informs the decision-making processes, promoting more effective collaboration and investment strategies in the evolving landscape of scientific research and education. Future work may incorporate advanced language models and interpretable frameworks, enabling more refined tasks such as causal knowledge reasoning, knowledge evolution analysis, structured knowledge retrieval, and personalized academic exploration.

\section*{Data and materials availability}
The code and data for reproducing our main results are publicly available at: \href{https://github.com/xiangshb/Knowledge-Navigation}{https://github.com/xiangshb/Knowledge-Navigation}.

\section*{Acknowledgments}
The computation in this study was supported by the Center for Computational Science and Engineering of SUSTech.

\section*{Funding}
This work was supported by the National Natural Science Foundation of China [NSFC62006109, NSFC12031005]; the 13th Five-year Plan for Education Science Funding of Guangdong Province [2020GXJK457]; the Stable Support Plan Program of Shenzhen Natural Science Fund under [20220814165010001]; and the Major Key Project of PCL [PCL2023A09].

\section*{Competing interests}
The authors declare that they have no known competing financial interests or personal relationships that could have appeared to influence the work reported in this paper.

\printcredits

\bibliographystyle{unsrt}

\bibliography{refs}




\appendix
\renewcommand{\thefigure}{S\arabic{figure}}
\setcounter{figure}{0}
\renewcommand{\thetable}{S\arabic{table}}
\setcounter{table}{0}
\twocolumn[
    \begin{@twocolumnfalse}
        \begin{center}
     {\LARGE Supplementary Information \par}
     \vspace{1cm} 
        \end{center}
    \end{@twocolumnfalse}
]

\section{Materials and Methods}
\subsection{Dataset description}
\begin{figure*}[!ht]
    \centering
     \begin{minipage}{0.46\textwidth}
         \centering
         \begin{subfigure}[b]{\linewidth}
     \centering
     \includegraphics[width=\linewidth]{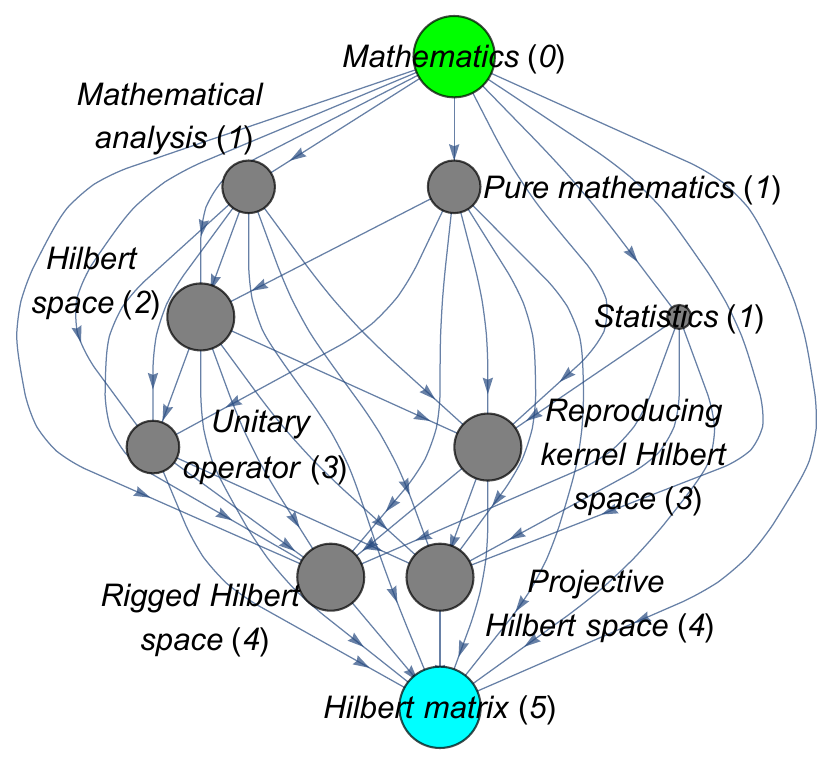}
     \caption{Sub-concept tree between Math. and Hilbert matrix}
     \label{concept_tree_math}
         \end{subfigure}
     \end{minipage}
    \begin{minipage}{0.46\textwidth}
        \centering
        \begin{subfigure}[b]{\linewidth}
     \centering
     \includegraphics[width=\linewidth]{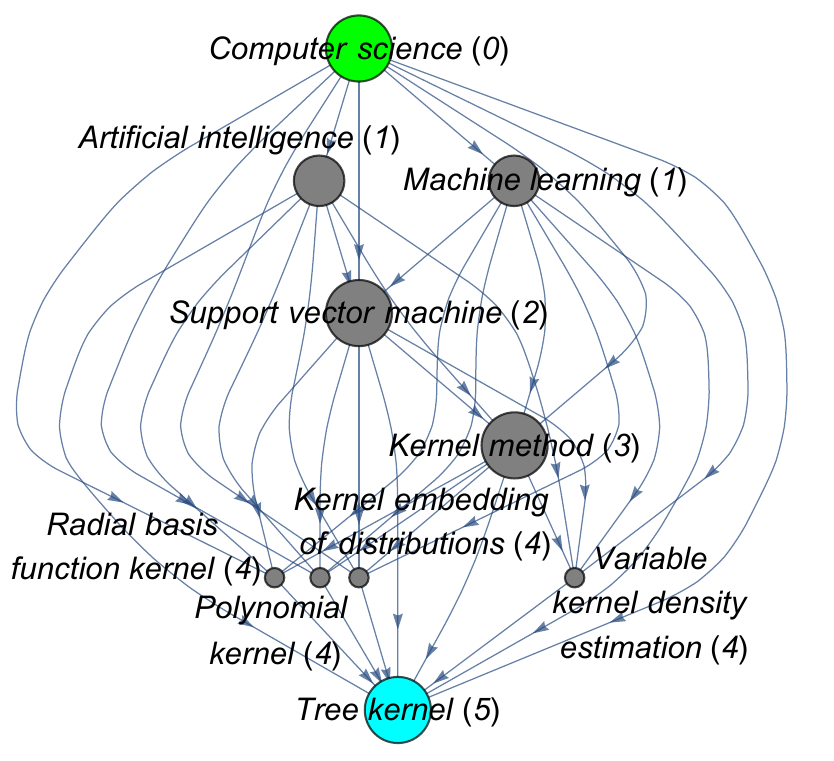}
     \caption{Sub-concept tree between C.S. and Tree kernel}
     \label{concept_tree_cs}
        \end{subfigure}
    \end{minipage}
     \begin{minipage}{0.46\textwidth}
         \centering
         \begin{subfigure}[b]{\linewidth}
     \centering
     \includegraphics[width=\linewidth]{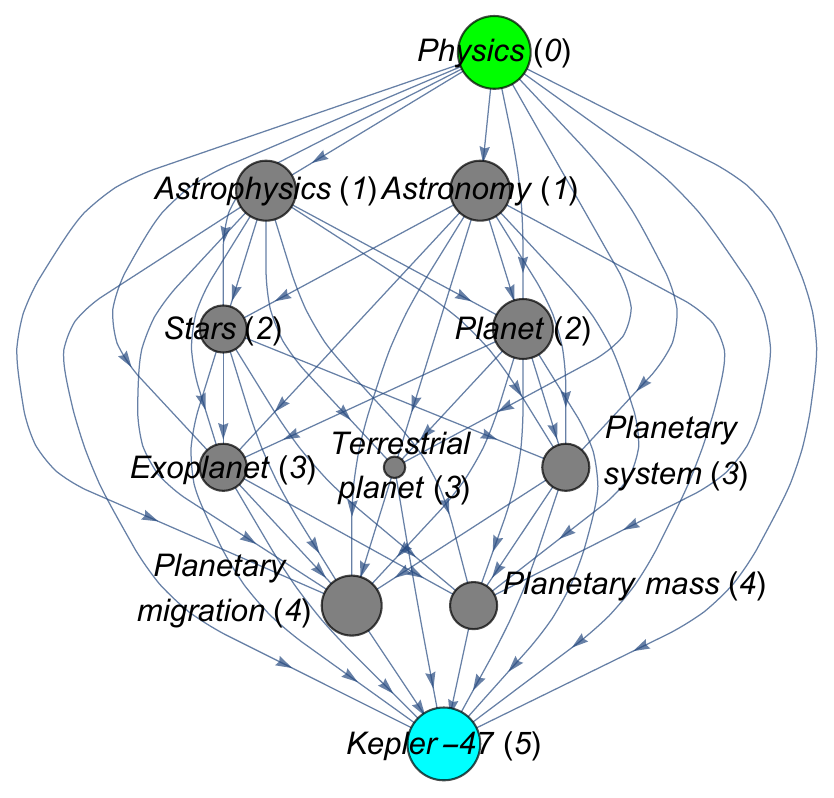}
     \caption{Sub-concept tree between Phys. and Kepler-47}
     \label{concept_tree_physics}
         \end{subfigure}
     \end{minipage}
    \begin{minipage}{0.46\textwidth}
        \centering
        \begin{subfigure}[b]{\linewidth}
     \centering
     \includegraphics[width=\linewidth]{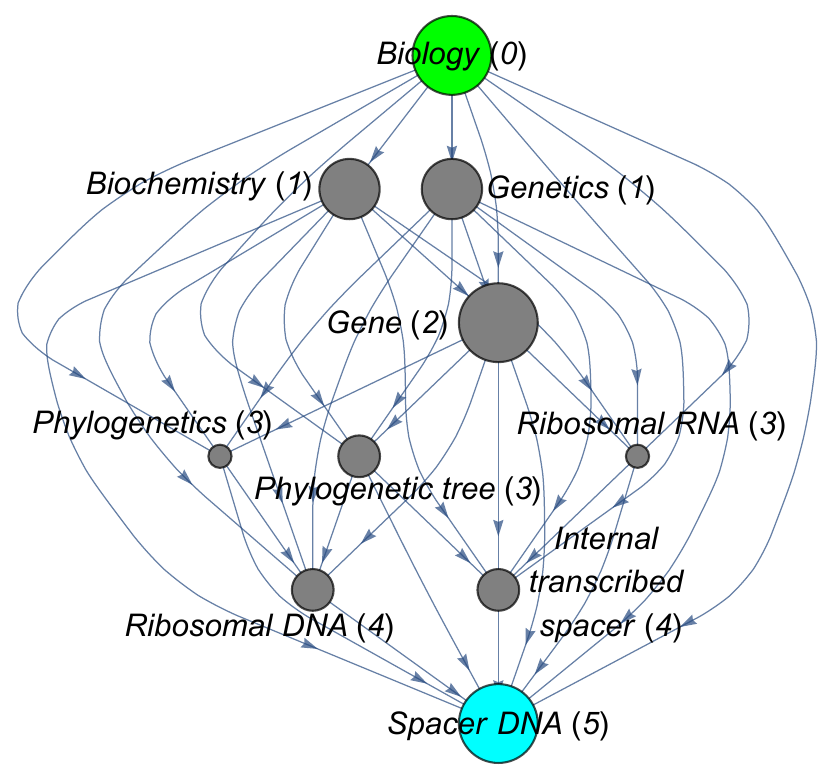}
     \caption{Sub-concept tree between Bio. and Spacer DNA}
     \label{concept_tree_biology}
        \end{subfigure}
    \end{minipage}
    \caption{\textbf{Examples sub-concept tree in OpenAlex dataset}. The size of the nodes corresponds to the degree of the nodes. Each directed edge in the graph represents the ancestral relationship between the ancestor and descendant concept. The number within the parentheses following the concept name signifies the concept's level. The hierarchy of the concept tree starts from level 0 at the top and gradually descends to level 5 at the bottom. (a) Sub-concept tree rooting from discipline Mathematics to level 5 concept Hilbert matrix. (b) Sub-concept tree rooting from discipline Computer science to level 5 concept Tree kernel. (c) Sub-concept tree rooting from discipline Physics to level 5 concept Kepler 47. (d) Sub-concept tree rooting from discipline Biology to level 5 concept Spacer DNA. }
    \label{examples_of_sub_concept_tree}
\end{figure*}

In our study, we utilized the OpenAlex dataset, which organizes concepts hierarchically across six levels (from level 0 to level 5). Each publication within this dataset is linked to multiple concepts, spanning levels 0 through 5. The dataset encompasses approximately 65,000 concepts in total. For a focused and comprehensive analysis, we selected the top 20,000 concepts based on the highest number of associated works. At the foundation of the hierarchy are 19 level 0 concepts, each representing a distinct academic discipline. Illustrative examples of these sub-concept trees are provided in Fig. \ref{examples_of_sub_concept_tree}, each originating from a distinct discipline concept and extending to level 5 concepts. Specifically, the trees start from the disciplines of Mathematics, Computer Science,  Physics, and Biology, connecting to their respective level 5 concepts: Hilbert matrix, Tree kernel, Kepler 47, and Spacer DNA. 

Taking Fig. \ref{concept_tree_math} as an illustration, we observe multiple directed graph paths bridging the discipline of Mathematics with the concept of Hilbert matrix. One such path unfolds as follows: [Mathematics $\rightarrow$ Mathematical analysis $\rightarrow$ Hilbert space $\rightarrow$ Unitary operator $\rightarrow$ Hilbert matrix]. This and similar sub-concept trees delineate the ancestral structure, shedding light on the underlying structure of knowledge within each discipline.

\begin{table*}[]
    \centering
    \caption{Abbreviations of 19 disciplines.}
    \begin{tabular}{c|l|l|c|l|l|c|l|l}
        \hline
        ID & \multicolumn{1}{c|}{\textbf{Discipline}} & \multicolumn{1}{c|}{\textbf{Abbr.}} & ID & \multicolumn{1}{c|}{\textbf{Discipline}} & \multicolumn{1}{c|}{\textbf{Abbr.}} & ID & \multicolumn{1}{c|}{\textbf{Discipline}} & \multicolumn{1}{c}{\textbf{Abbr.}} \\ \hline
        1  & Art   & Art  & 8  & Environmental science    & Env.Sci.    & 14 & Medicine      & Med.       \\ \hline
        2  & Biology   & Bio.  & 9  & Geography     & Geog.  & 15 & Philosophy    & Philo.     \\ \hline
        3  & Business      & Busi.    & 10 & Geology   & Geol.  & 16 & Physics   & Phys.      \\ \hline
        4  & Chemistry     & Chem.    & 11 & History   & Hist.  & 17 & Political science   & Pol.Sci.   \\ \hline
        5  & Computer science    & C.S.  & 12 & Materials science   & Mat.Sci. & 18 & Psychology    & Psych.     \\ \hline
        6  & Economics     & Econ.  & 13 & Mathematics   & Math. & 19 & Sociology     & Sociol.    \\ \hline
        7  & Engineering   & Eng.  &   &     &  &   &       &      \\ \hline
    \end{tabular}
    \label{discipline_abbreviations}
\end{table*}

\begin{table*}[]
    \centering
    \caption{Discipline group partition of 19 disciplines}
    \begin{tabular}{l|l|l|l}
        \hline
        \multicolumn{1}{c|}{\textbf{Functional group}} & \multicolumn{1}{c|}{\textbf{Disciplines}} & \multicolumn{1}{c|}{\textbf{Functional group}} & \multicolumn{1}{c}{\textbf{Disciplines}}  \\ \hline
        Theoretical    & Mathematics, Physics   & Societal    & Sociology, Political science, Psychology  \\ \hline
        Applied        & Computer science, Engineering & Economic   & Economics, Business      \\ \hline
        Chemical       & Chemistry, Materials science    & Humanities  & Philosophy, History, Art    \\ \hline
        Biomedical     & Biology, Medicine  & Geographical   & Geography, Geology, Environmental science \\ \hline
    \end{tabular}
    \label{discipline_groups}
\end{table*}

\subsection{Concept discipline category classification}
We denoted the set of 19 level 0 concepts as $C^0$, which are shown in   Table \ref{discipline_abbreviations} from No.1 to No.19. To further refine our analysis, we introduced three additional categories: ``Disciplinary'', ``Interdisciplinary'', and ``Multi-interdisciplinary''. These categories help streamline the classification process and the specific scope of concepts under each category will be elucidated in subsequent sections.

For concepts that are not at level 0, denoted as $c\in \overline{C^0}$ and ranging from level 1 to level 5 (as shown in Fig. \ref{concept_tree_math} to \ref{concept_tree_biology}), there is no direct information to tell which discipline (defined by level 0 ancestor roots in $C^0$) this concept mainly belongs to, it's natural that we consider its root ancestors, the level 0 concepts in $C^0$ to determine its main discipline. Let AR$_c\subseteq C^0$ be the set of ancestor roots of concept $c\in \overline{C^0}$, and denote $L_c \in \text{AR}_c$ as the classified discipline label of concept $c$, hence we determine the discipline $L_c$ as the main discipline that concept $c$ belongs to. Firstly, we remove the concepts with no direct discipline, and subsequently classified remaining concepts into two groups $S$ and $M$.

$S=\left\{c\left | c\in \overline{C^0}\right .,\left| \text{AR}_c \right|=1 \right\}$, a total of 17,508 concepts with a single AR. This single AR of each concept $c$ is the discipline category of $c$, without causing any ambiguity.

$M=\left\{c\left | c\in \overline{C^0}\right .,\left| \text{AR}_c \right|>1 \right\}$, a total of 47,002 concepts with multiple AR. We determine the primary discipline $L_c$ for each concept $c$ by analyzing the graph paths between its multiple ancestor roots (AR) and the concept itself. Let $P (d \rightarrow c)$ represent the number of distinct paths from a discipline $d$ to concept $c$ in the concept graph. As illustrated in Fig. \ref{concept_tree_math}, multiple paths typically exist from the root concept (e.g., Mathematics) at the top to a specific concept (e.g., Hilbert matrix) at the bottom of the sub-tree. We determine the main discipline $L_c$ of concept $c$ as the ancestor root that maximizes the number of graph paths to $c$ that
\begin{equation}
    L_c = arg\underset{d\in \text{AR}_c}{\max} P\left( d\rightarrow c \right).
    \label{discipline_argmax}
\end{equation}
The concept $c \in M$ is deemed classifiable if there exists a unique $L_c$ that exhibits the maximum number of graph paths to $c$. Conversely, $c$ is considered indistinguishable if multiple $L_c \in \text{AR}_c$ share the same maximum number of graph paths to $c$.
The majority of concepts in $M_2$ are classifiable. We define the extended classifiable concepts $S^+$ as the union of concepts in $S$ and classifiable concepts in $M$, comprising 49,275 concepts, which we term ``Disciplinary'' concepts. The remaining indistinguishable concepts in $M$, denoted as $M^-$ and termed ``Multi-disciplinary'' concepts, account for 15,235 concepts.

Overall, this method allows us to systematically assign a discipline category to all the concepts under consideration, ensuring each is accurately categorized based on its most prominent connections within the knowledge network.

\section{Extended results}
\subsection{Discipline propensity of concepts from different disciplines}
\begin{figure*}[h]
    \centering
    \includegraphics[width=\textwidth]{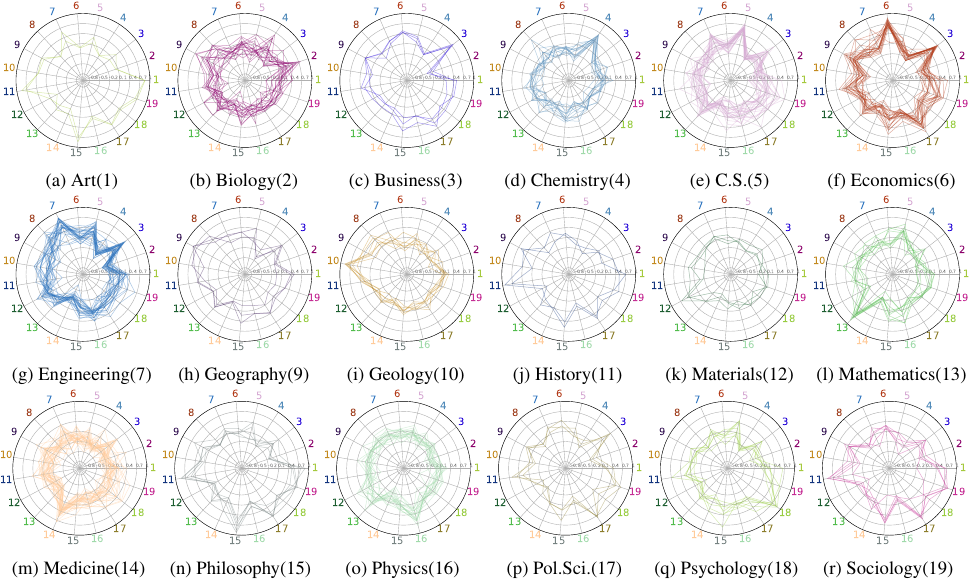}
    \caption{Discipline similarity radar rapping: level 1 sub-concepts. Each radar map, numbered from (1) to (19) excluding the Env.Sci (8), illustrates the propensity of level 1 concepts from a specific discipline towards the 18 disciplines.}
    \label{radar_mappings_all_disciplines_level_1}
\end{figure*}

In addition to the examples in the main text, we show more examples for other disciplines to elucidate the discipline propensity of concepts, we employ radar maps to visualize the inclination of level 1 and level 2 concepts towards their respective labeled disciplines. These visual representations are provided in Figs. \ref{radar_mappings_all_disciplines_level_1} and \ref{radar_mappings_all_disciplines_level_1}, offering insights into the domain tendencies of disciplinary concepts across various fields.

Fig. \ref{radar_mappings_all_disciplines_level_1} presents a series of radar maps for 19 disciplines listed in Table \ref{discipline_abbreviations}, each radar map represents the discipline propensity of level 1 concepts from a distinct discipline to all 19 disciplines. In these radar maps, the most outward-pointing angle (MOPA)  towards the labeled discipline indicates that level 1 concepts exhibit a notably higher cosine similarity with their labeled discipline in comparison to other disciplines. Similarly, Fig. \ref{radar_mappings_all_disciplines_level_1} displays a set of radar maps of discipline propensity of level 2 concepts of each discipline against 19 disciplines respectively. Consistently with level 1 concepts, the MOPA in the radar map of each discipline towards this discipline underscores that level 2 concepts maintain a higher cosine similarity with their labeled discipline as opposed to other disciplines.

\begin{figure*}[h]
    \centering
    \includegraphics[width=\textwidth]{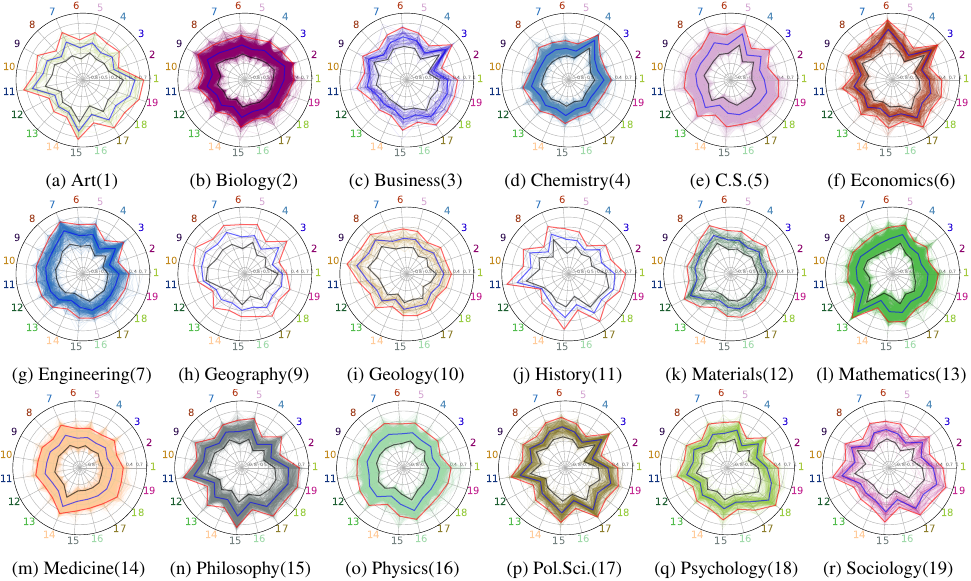}
    \caption{Discipline similarity radar rapping: level 2 sub-concepts. Each radar map, numbered from (1) to (19) excluding the Env.Sci (8), illustrates the propensity of level 2 concepts from a specific discipline towards the 18 disciplines.}
    \label{radar_mappings_all_disciplines_level_2}
\end{figure*}

Figures \ref{radar_mappings_all_disciplines_level_1} and \ref{radar_mappings_all_disciplines_level_2} present radar charts that depict the discipline propensity of level 1 and level 2 sub-concepts across 18 disciplines. Discipline Env.Sci (8) is excluded due to the limited number of concepts, as specified in Table \ref{discipline_abbreviations}. Each radar map displays the measured propensity of sub-concepts from a specific discipline towards all 19 disciplines. In both figures, each radar map contains 19 axes corresponding to the disciplines, with the outward-pointing degree of each axis representing the calculated propensity of the sub-concepts towards that particular discipline. For instance, in the Mathematics radar map, the most outward-pointing angle (MOPA) aligns with the Mathematics axis, indicating that Mathematics sub-concepts have the highest calculated propensity towards their own discipline compared to the other 18 disciplines. This pattern is consistent across all disciplines and both hierarchical levels, with each exhibiting the strongest propensity towards its labeled discipline. The trend observed in Fig. \ref{radar_mappings_all_disciplines_level_1} for level 1 sub-concepts is mirrored in Fig. \ref{radar_mappings_all_disciplines_level_2} for level 2 sub-concepts. These visualizations effectively illustrate the strong association between sub-concepts and their respective labeled disciplines across both hierarchical levels, as determined by our calculations, highlighting the distinctiveness and cohesiveness of disciplinary knowledge structures.

\subsection{Examples of analogy inference networks}
In Table \ref{more_analogy_inference}, we present a series of concept analogy inference examples. For each example, starting with a seed concept \( a \) and two analogous concepts \( c \) and \( d \), we perform a two-step analogy inference process. In the first step, we derive the initial inference results \( b^{+} \)and \( b^{-} \)from the seed concept \( a \). Using these initial results as new seed concepts, we then proceed to the second step of inference. The outcomes of the second step of inference \( b^{++} \)and \( b^{+-} \)derived from \( b^{+} \), and \( b^{-+} \)and \( b^{--} \) derived from \( b^{-} \) are systematically documented in each row of Table \ref{more_analogy_inference}, respectively.

\begin{table*}[h]
    \centering
    \caption{Two-step analogy inference cases in scientific concepts}
    \begin{tabular}{l|l|l}
        \hline
        \multicolumn{1}{c|}{Case} & \multicolumn{1}{c|}{No. 1} & \multicolumn{1}{c}{No. 2} \\ \hline
        $a$         & Distributed computing (C.S.) & Combinatorics (Math.)  \\ \hline
        $c$         & Computer graphics (C.S.) & Information retrieval (C.S.)  \\ \hline
        $d$         & Statistics (Math.) & Statistics (Math.)  \\ \hline
        $b^+$     & Erasure (C.S.,Math.) & Noncentral chi-squared distribution (Bio.,Chem.,Math.,Med.)  \\ \hline
        $b^{++}$  & Binary symmetric channel (C.S.,Eng.,Math.) & Logistic distribution (C.S.,Math.,Med.)  \\ \hline
        $b^{+-}$  & Emerging technologies (C.S.Mat.Sci.) & Negation (C.S.,Philo.)  \\ \hline
        $b^-$     & Emerging technologies (C.S.Mat.Sci.) & Ramsey's theorem (C. S.,Math.)  \\ \hline
        $b^{-+}$  & Computational topology (Chem.,Math.,Phys.) & Conjunctive query (C.S.)  \\ \hline
        $b^{--}$  & Desk (C.S.,Eng.,Pol.Sci.) & Noncentral chi-squared distribution (Bio.,Chem.,Math.,Med.)  \\ \hline
        \multicolumn{1}{c|}{Case} & \multicolumn{1}{c|}{No. 3} & \multicolumn{1}{c}{No. 4} \\ \hline
        $a$         & Database (C.S.) & Data science (C.S.) \\ \hline
        $c$         & Computer network(C.S.) & Computer network (C.S.) \\ \hline
        $d$         & Mathematical analysis (Math.) & Mathematical analysis (Math.) \\ \hline
        $b^+$     & Heat equation (Math.,Phys.) & Heat equation (Math.,Phys.) \\ \hline
        $b^{++}$  & Boundary value problem (Math.,Phys.) & Boundary value problem (Math.,Phys.) \\ \hline
        $b^{+-}$  & Live migration (C.S.,Pol.Sci.) & Live migration (C.S.,Pol.Sci.) \\ \hline
        $b^-$     & Virtualization (C.S.,Pol.Sci.) & Cloud testing (C.S.,Pol.Sci.) \\ \hline
        $b^{-+}$  & Boundary value problem (Math.,Phys.) & Elementary theory (C.S.,Math.) \\ \hline
        $b^{--}$  & Server (C.S.) & Virtualization (C.S.,Pol.Sci.) \\ \hline
        \multicolumn{1}{c|}{Case} & \multicolumn{1}{c|}{No. 5} & \multicolumn{1}{c}{No. 6} \\ \hline
        $a$         & Combinatorics (Math.) & Applied mathematics (Math.) \\ \hline
        $c$         & Multimedia (C.S.) & Computer graphics (C.S.) \\ \hline
        $d$         & Applied mathematics (Math.) & Statistics (Math.) \\ \hline
        $b^+$     & Skew-symmetric matrix (Math.,Phys.) & Bivariate analysis (C.S,,Math.) \\ \hline
        $b^{++}$  & Matrix differential equation (Math.,Phys.) & Logistic distribution (C.S.,Math:,Med.) \\ \hline
        $b^{+-}$  & Interactive video (C.S.) & Ray casting (C.S.,Eng.) \\ \hline
        $b^-$     & Two-way communication (C.S.,Eng.) & Critical system (Busi.,C.S.,Econ.,Eng.,Math.,Med.) \\ \hline
        $b^{-+}$  & Homoclinic bifurcation (Eng.,Math.,Phys.) & Odds (C.S.,Math.,Med.) \\ \hline
        $b^{--}$  & Critical system (Busi.,C.S.,Econ.,Eng.,Math.,Med.) & Computer Science and Engineering (C.S.,Eng.) \\ \hline
    \end{tabular}
    \label{more_analogy_inference}
\end{table*}

\begin{figure*}[htbp]
    \centering
    \begin{subfigure}[b]{\textwidth}
        \centering
        \includegraphics[width=\textwidth]{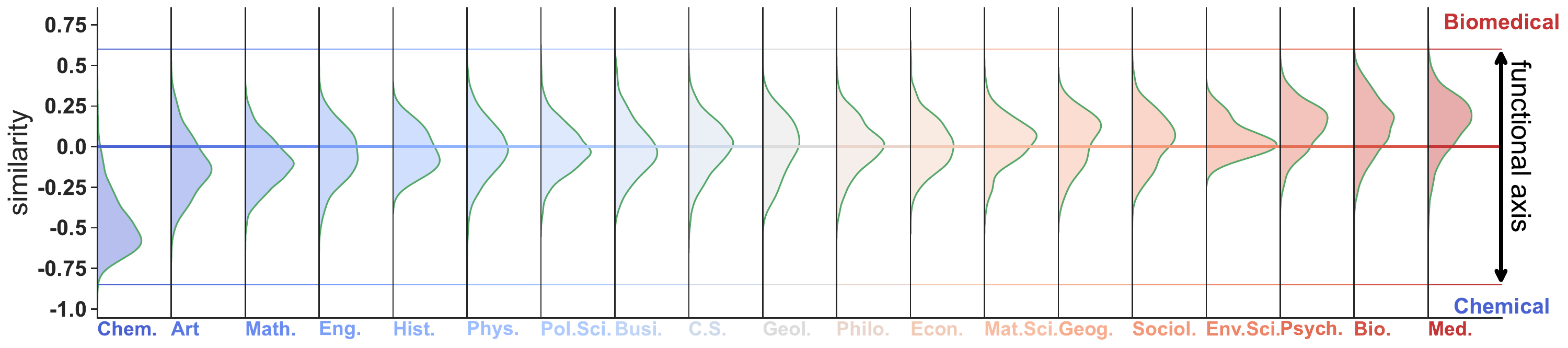}
        \caption{Functional axis between Chemical and Biomedical}
        \label{functional_axis_chemical_biomedical}
    \end{subfigure}
    \begin{subfigure}[b]{\textwidth}
        \centering
        \includegraphics[width=\textwidth]{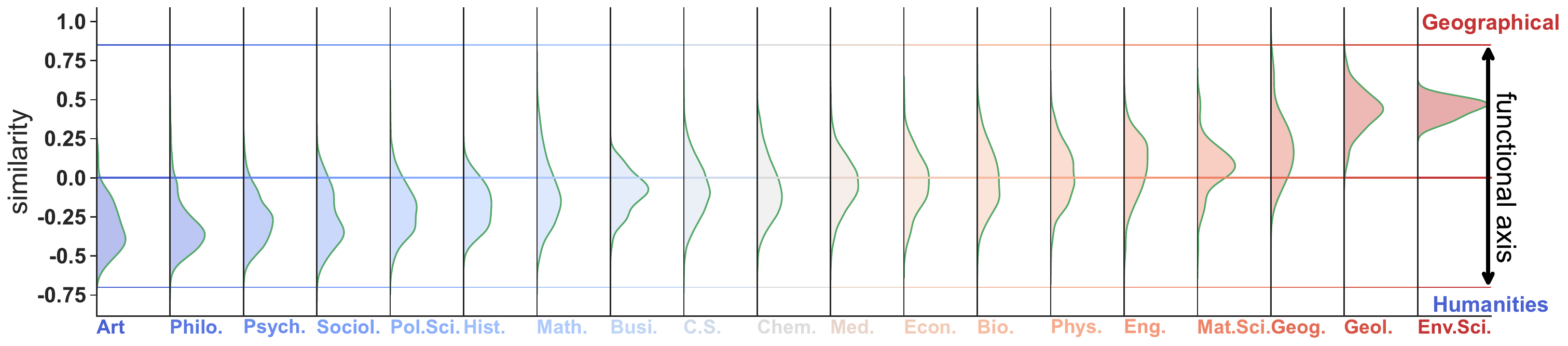}
        \caption{Functional axis between Humanities and Geographical}
        \label{functional_axis_humanities_geographical}
    \end{subfigure}
    \caption{\textbf{Functionality projections of knowledge from 19 disciplines on predefined axes}. (a) Projection of all disciplines on functional axis between Chemical and Biomedical. (b) Projection of all disciplines on functional axis between Humanities and Geographical.}
    \label{functionality_projections_SI}
\end{figure*}

\subsection{Examples of concept pathways}
In Table \ref{more_concept_pathways}, we detail additional examples of concept pathways to illustrate the interconnectedness within the knowledge network. For instance, consider the pathway delineated between the concepts ``Statistics'' and ``Artificial Intelligence,'' presented as the second example in the table. This pathway includes an intermediary concept: ``Weighting.'' The presence of ``Weighting'' as a pivotal node is notably fitting, given that contemporary AI methodologies heavily emphasize network weight learning. This focus is instrumental in achieving precise mappings, whether they are one-to-one or many-to-many, which are crucial for enhancing accuracy in AI applications.

The provided examples illustrate the shortest semantic paths between pairs of concepts using word2vec vector representations, with training data sorted by the publication time of scientific papers. These paths reveal the semantic and logical relationships between concepts, offering insights into their interconnectedness and practical applications. For instance, the path from ``Statistics'' to ``Sociology'' includes ``Categorical variable'' and ``Relation (database),'' indicating the importance of statistical methods in analyzing sociological data. Similarly, the path from ``Artificial Intelligence'' to ``Bioinformatics'' via ``Interpretability'' emphasizes the need for transparency in AI applications within bioinformatics. Each path demonstrates domain-specific associations that are logical and coherent, such as the path from ``Statistics'' to ``Plant growth,'' which includes ``Best linear unbiased prediction'' and ``Diallel cross,'' reflecting the critical role of statistical methods in agricultural research and plant breeding.

These paths underscore the interdisciplinary nature of modern research and can guide researchers in identifying key intermediate concepts and methodologies that bridge different fields. For example, the connection between ``Artificial Intelligence'' and ``Sociology'' via ``Representation (politics)'' and ``Meaning (existential)'' suggests that AI's impact on societal structures and existential questions is a critical area for future research. Additionally, the path from ``Pure mathematics'' to ``Driving simulator'' via ``Optimal control'' and ``Control (management)'' indicates the application of mathematical optimization and control theory in developing advanced driving simulations. By revealing logical connections between seemingly disparate fields, these paths can foster cross-disciplinary collaboration. For instance, the path from ``Green building'' to ``Human genetics'' through ``Energy engineering'' and ``Gene cluster'' suggests potential interdisciplinary research opportunities in sustainable building practices and genetic studies. In summary, these shortest semantic paths provide a structured and meaningful way to understand the relationships between concepts across different domains, offering valuable insights for guiding research directions and enhancing cross-disciplinary collaboration.

\begin{table*}[]
    \centering
    \caption{Examples of concept pathways between interested concept pairs}
    \begin{tabular}{c|l}
        \hline
        No. & \multicolumn{1}{c}{Shortest concept pathway between two concepts} \\ \hline
        1  & [Statistics, Categorical variable, Relation (database), Meaning (existential), Sociology] \\ \hline
        2  & [Statistics, Weighting, Artificial intelligence] \\ \hline
        3  & [Statistics, Probability distribution, Random walk, Complex system] \\ \hline
        4  & [Statistics, Best linear unbiased prediction, Diallel cross, Cultivar, Phaseolus, Germination, Plant growth]  \\ \hline
        5  & [Artificial intelligence, Representation (politics), Meaning (existential), Sociology] \\ \hline
        6  & [Artificial intelligence, Representation (politics), Semiotics, Dialectic, Pragmatism, Social science] \\ \hline
        7  & [Artificial intelligence, Information processing, Judgement, CLARITY, Need to know, Nursing] \\ \hline
        8  & [Artificial intelligence, Segmentation, Lung biopsy, Bronchoscopy, Cardiothoracic surgery, Vascular surgery] \\ \hline
        9  & [Artificial intelligence, Task (project management), Driving simulator] \\ \hline
        10  & [Artificial intelligence, Interpretability, Bioinformatics] \\ \hline
        11  & [Pure mathematics, Discrete mathematics, Graph, Biological network, Bioinformatics] \\ \hline
        12  & [Applied mathematics, Convergence (economics), Optimal control, Control (management), Driving simulator] \\ \hline
        13  & [Combinatorics, Upper and lower bounds, Channel capacity, Fading, Multipath propagation, Radar] \\ \hline
        14  & [Green building, ASHRAE 90.1, Natural ventilation, Airflow, Venturi effect, Blood viscosity] \\ \hline
        15  & [Green building, Energy engineering, Energy resources, Gene cluster, Genetics, Human genetics] \\ \hline
    \end{tabular}
    \label{more_concept_pathways}
\end{table*}

\begin{figure*}[!ht]
    \centering
     \begin{minipage}{0.46\textwidth}
         \centering
         \begin{subfigure}[b]{\linewidth}
     \centering
     \includegraphics[width=\linewidth]{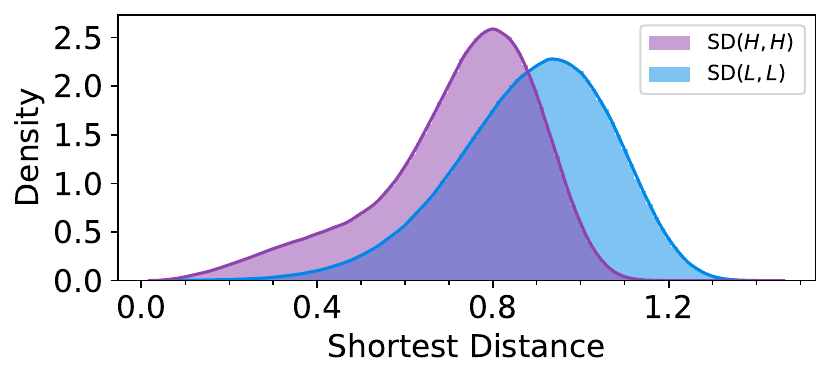}
     \caption{Shortest distance within $H$ vs. $L$ concepts by closeness}
     \label{closeness_SD_pdf}
         \end{subfigure}
     \end{minipage}
     \begin{minipage}{0.46\textwidth}
         \centering
         \begin{subfigure}[b]{\linewidth}
     \centering
     \includegraphics[width=\linewidth]{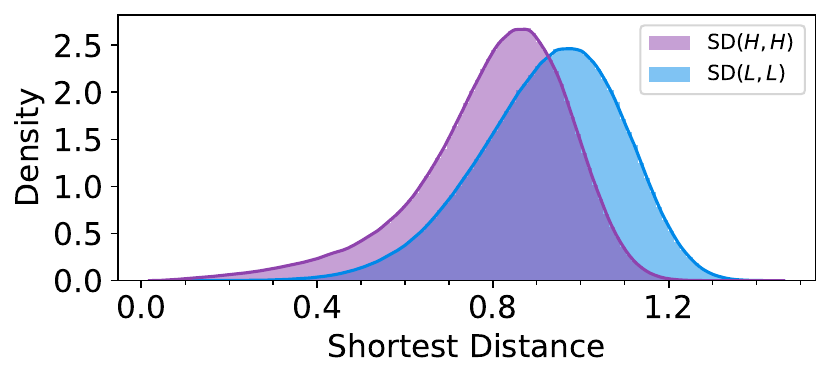}
     \caption{Shortest distance within $H$ vs. $L$ concepts by betweenness}
     \label{betweenness_SD_pdf}
         \end{subfigure}
     \end{minipage}
    \caption{\textbf{Shortest distance between concepts with different centrality}. (a) Comparison of shortest distance distribution between high closeness centrality concepts vs. low closeness centrality concepts. (b) Comparison of shortest distance distribution between high betweenness centrality concepts vs. low betweenness centrality concepts.}
    \label{SD_pdf_of_centrality}
\end{figure*}

\begin{figure*}[h]
    \centering
    \begin{subfigure}[m]{0.48\textwidth}
        \centering
        \includegraphics[width=\linewidth]{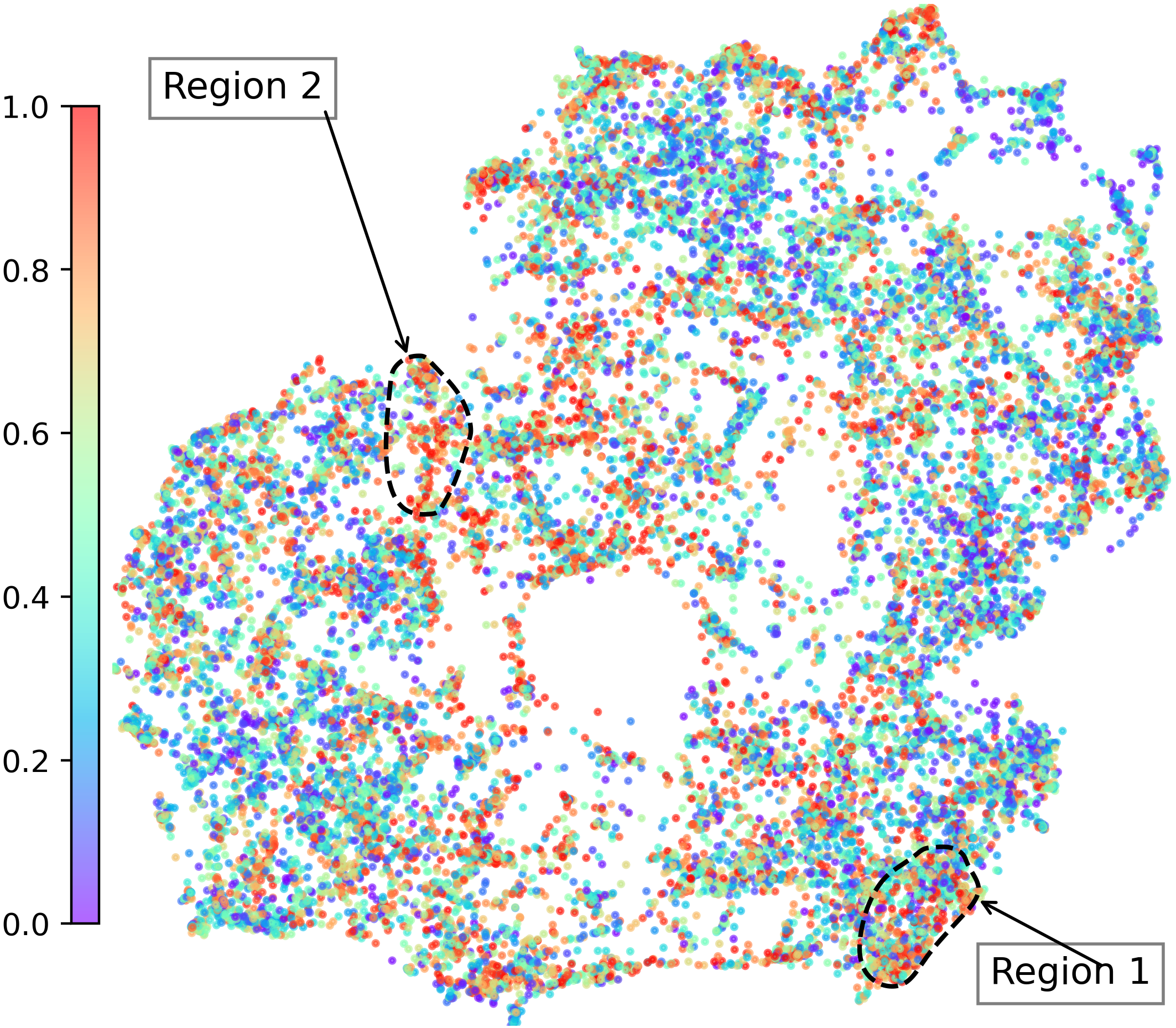}
        \caption{Heatmap of concept betweenness centrality}
        \label{betweenness_centrality}
    \end{subfigure}
    \begin{subfigure}[m]{0.45\textwidth}
        \centering
        \begin{subfigure}{\textwidth}
     \centering
     \includegraphics[width=\linewidth]{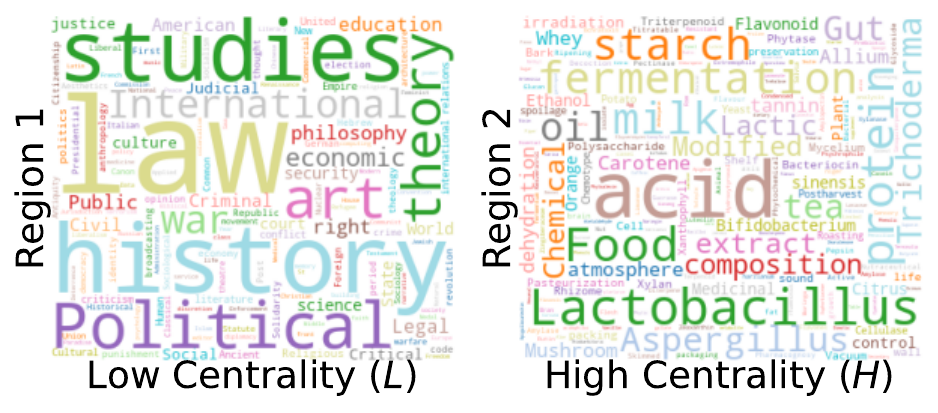}
        \end{subfigure}
        \begin{subfigure}{\textwidth}
     \centering
     \includegraphics[width=\linewidth]{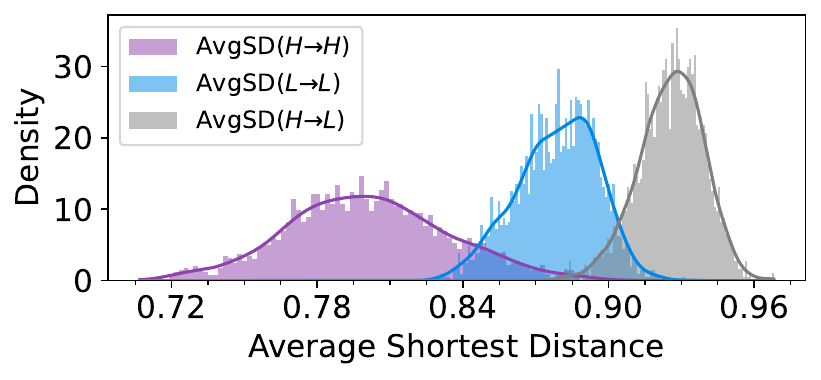}
     \caption{AvgSD between 2k $H$ and 2k $L$ concepts by betweenness}
     \label{ave_betweenness_SD_pdf}
        \end{subfigure}
    \end{subfigure}
    \caption{\textbf{Centrality measure the importance of interdisciplinary concepts}. (a) Embedding map of 20,000 key concepts, color-coded by betweenness centrality. (b) Concept word clouds of two selected regions, region 1 mainly contains concepts with low betweenness centrality, and region 2 mainly contains concepts with high betweenness centrality.}
    \label{betweenness_centralities}
\end{figure*}

\end{document}